\documentclass[twocolumn,aps,prb,reprint,groupedaddress,superscriptaddress]{revtex4}
\usepackage{amsmath}
\usepackage{amssymb}
\usepackage{etoolbox}
\usepackage{graphicx}
\usepackage{color}

\newcommand{\be}{\begin{equation}}
\newcommand{\ee}{\end{equation}}
\def\ba{\begin{aligned}}
\def\ea{\end{aligned}}

\newcommand{\bea}{\begin{eqnarray}}
\newcommand{\eea}{\end{eqnarray}}

\begin{document}

\title{A random matrix model with localization and ergodic transitions.}
\author{V.~E.~Kravtsov}
 \affiliation{Abdus Salam International Center for Theoretical Physics, Strada Costiera 11, 34151 Trieste, Italy}
 \affiliation{L. D. Landau Institute for Theoretical Physics, Chernogolovka, Russia}
\author{ I.~M.~Khaymovich}
 \email{ivan.khaymovich@aalto.fi}
 \affiliation{Low Temperature Laboratory, Department of Applied Physics, Aalto University, FI-00076 Aalto, Finland}
 \affiliation{Institute for Physics of Microstructures, Russian Academy of Sciences, 603950 Nizhny Novgorod, GSP-105, Russia }
\author{E.~Cuevas}
 \affiliation{Departamento de F\'{\i}sica, Universidad de Murcia, E30071 Murcia, Spain}
\author{M. Amini}
 \affiliation{Department of Physics, University of Isfahan(UI), Hezar Jerib, 81746-73441, Isfahan, Iran}

\begin{abstract}
Motivated by the problem of Many-Body Localization and the recent numerical results for the level and eigenfunction
statistics on the random regular graphs, a generalization of the Rosenzweig-Porter random matrix model is suggested that possesses two
transitions. One of them is the Anderson localization transition
from the localized to the extended states. The other one is the ergodic  transition from the
extended non-ergodic (multifractal) states to the extended ergodic states.
We confirm the existence of both transitions by computing the two-level spectral correlation function, the spectrum of multifractality $f(\alpha)$ and the wave function overlap which consistently demonstrate these two transitions.
\end{abstract}


\maketitle

\section{Introduction}
Motivated by
the problem of Many-Body (MB) Localization \cite{BAA} and the applicability of the Boltzmann's statistics in interacting disordered media \cite{Huse}, there was recently a revival of
interest to the Anderson localization (AL) problem on hierarchical lattices such as the Bethe lattice (BL) or the random
regular graph (RRG). Due to hierarchical structure of the Fock space connected by the two-body interaction,
statistics of random wave functions in such models is an important playground for MB localization.
In particular, the non-ergodic extended phase on disordered hierarchical lattices could model a breakdown of conventional Boltzmann statistics
in interacting MB systems and an emergence of a phase of a ``bad metal'' \cite{AltshulerJoffe} or unconventional fluid phases \cite{AltshulerShlyapnikov} in systems of interacting particles.

However, even for the one-particle AL existence of such a phase in a finite interval of disorder strengths is a highly non-trivial issue.

According to earlier studies \cite{AbouChac-And,Mir-Fyod-sparse} there is only one transition in such models
at a disorder strength $W=W_{AT}$ which is the AL transition that separates the localized and ergodic extended states. However, recent numerical studies \cite{Biroli} of level statistics on
 RRG seem to indicate on the second transition at $W=W_{ET}<W_{AT}$ which is identified as the transition between the ergodic and non-ergodic extended states.
Subsequent studies \cite{Our-BL,Biroli-Levy} raise doubts about the existence of the second transition on RRG. Numerical results of Ref.~\cite{Our-BL} indicate on the non-ergodic states on RRG in a wide range of disorder strengths down to very low disorder $W=5\ll W_{AT}\approx 17.5$, while in Ref.~\cite{Biroli-Levy}
it is demonstrated how an apparent non-ergodic behavior for the intermediate matrix sizes N in Levy Random Matrix (RM) ensemble evolves into the ergodic one at larger $N$'s.
Complexity of RRG and the controversy associated with existence of the  ergodic transition at $W=W_{ET}$ necessitate a search for a simpler model in which such a transition may occur.

Inspired by the success of Wigner-Dyson RM theory \cite{Mehta} which predictions are relevant in such seemingly different fields of physics as nuclear physics and nano- and mesoscipic physics, our goal is to search for a RM model that
would be able to give a simple and universal description of all the three phases: good metal, MB insulator and ``bad metal'', which are relevant in the problem of MB localization.
An important heuristic argument to construct such a model is that RRG with disordered on-site energies $\varepsilon_{i}$ is
essentially a two-step disorder ensemble. The disorder of the first level is the structural disorder due to the random structure of RRG
where each of $N$ sites of the graph
is connected with the fixed number $K+1$ of other sites in a random manner. An ensemble of tight-binding models on such graphs with deterministic on-site energies $\varepsilon_{i}$ and hopping integrals is believed to
be equivalent to the Gaussian RM ensemble \cite{Smiliansky}. The disorder of the second level is produced by randomization of
$\varepsilon_{i}$ fluctuating independently around zero with the distribution
function $p(\varepsilon)$. For numerical calculations this distribution is often taken in the form
$p(\varepsilon)=\frac{1}{W}\cdot\theta\left(\frac{W}{2}-|\varepsilon|\right)$, with $\theta(x)$ being the Heaviside step function.

One can expect that the following RM ensemble (the Rosenzweig-Porter (RP) ensemble \cite{RPort}) is a close
relative of RRG with on-site energy $\varepsilon_{i}$ disorder. It is an ensemble of $N\times N$ random Hermitian matrices
which entries $H_{nm}$ with $n > m$ are independent random Gaussian numbers, real for orthogonal RP model ($\beta=1$) and complex for the unitary RP model ($\beta=2$), fluctuating around zero with
the variance
$\langle |H_{nm}|^{2}\rangle =(\beta/2)\,\sigma$. The diagonal elements have the same properties with the variance
$\langle H_{nn}^{2}\rangle =1$. The case $\sigma=1$ corresponds to the Gaussian Orthogonal (GOE) or Gaussian Unitary (GUE) ensembles and represents
the structural disorder in RRG. The additional $\varepsilon_{i}$-disorder in RRG corresponds to $\sigma<1$.
One should also take into account that in order to significantly deviate from the GOE or GUE behavior, the ratio $\langle |H_{nm}|^{2}\rangle/\langle H_{nn}^{2}\rangle$ must be proportional to some negative power of the matrix size $N$, as the number of the off-diagonal terms is $\sim N$ times larger.
 Thus we consider the model:
\begin{gather}\label{model}
\langle H_{nn}^{2} \rangle = 1,\;\;\;\;\langle |H_{n\neq m}|^{2}\rangle = (\beta/2)\,\sigma = {\lambda^{2}}/{N^{\gamma}},
\end{gather}
where $\lambda\sim O(N^{0})$ and $\gamma$ is the main control parameter of the problem.
One can estimate the strength of disorder required for the Anderson localization transition as  corresponding
to the typical fluctuation of diagonal matrix element equal to the typical off-diagonal matrix element times the
coordination number $K$. For the coordination number $K\sim N$ (each site is connected with any other one)
this results in $\sqrt{\sigma} N \sim 1$, or $\sigma \sim 1/N^{2}$.   However, this
estimation does not take into account a random, sign-alternating character of the off-diagonal matrix elements.
It is likely that   for sign-alternating hopping there is another  relevant  coordination number  $\sim \sqrt{N}$
with the critical scaling   $\sigma\sim 1/N$. As we show below it corresponds to the ergodic transition.
 For technical reasons the most significant progress in the analytical studies of the model was achieved \cite{Pandey, BrezHik, ShapiroKunz} for the ``unitary''
RP (URP) ensemble. The conclusion was that at $\gamma=2$ the spectral form-factor (two-level correlation function) is neither of the Wigner-Dyson nor of the Poisson
form \cite{Pandey, Altland-Shapiro, ShapiroKunz} which is typical for the AL transition point.
In contrast, at $\gamma=1$ the level statistics was found to be  GUE \cite{BrezHik}.
The papers \cite{Pandey,BrezHik,ShapiroKunz} have a status of classic keynote papers in the field.

The Dyson ideas of the Brownian motion of energy levels first applied to the RP ensemble in Ref.~\cite{Pandey} were developed in the series of works \cite{Shukla2000, Shukla2005}. It was shown that the possible transitions in the level statistics are associated with the fixed points of parameter $\Lambda=\sigma(N)/[\delta(N)]^{2}\propto N^{-\gamma}/[\delta(N)]^{2}$, where $\delta(N)$ is the mean level spacing. Then assuming $\delta(N)\propto 1/N$ established in Ref.~\cite{ShapiroKunz} for $\gamma>1$ one obtains the transition point at $\gamma=2$. If, however, the Wigner-Dyson semicircle level density is assumed with $\delta(N)\propto 1/\sqrt{N}$, then the transition would occur at $\gamma=1$ \cite{Shukla2000}. Unfortunately, $\delta(N)\propto 1/\sqrt{N}$ only at $\gamma=0$. For $0<\gamma\leq 1$ the following result is valid (see e.g. Eq.~(170) in Ref.~\cite{RMTLect}) for the mean density of states $\rho(E)=\sqrt{2S-E^{2}}/(\pi S)$, where $S=\sum_{n}\langle |H_{nm}|^{2} \rangle \propto N^{1-\gamma}$. Thus we obtain $\delta(N)=1/(N\rho(0))\propto N^{-(1+\gamma)/2}$ for $0<\gamma\leq 1$ and $\delta(N)\propto 1/N$ for $\gamma>1$, resulting in $\Lambda\propto N$ for $0<\gamma\leq 1$ and $\Lambda\propto N^{2-\gamma}$ for $\gamma>1$. We conclude that the only fixed point of $\Lambda$ is possible at $\gamma=2$, and no transition at $\gamma=1$ can be obtained from the results of Refs.~\cite{Shukla2000,Shukla2005}.

In this paper by a more sophisticated analysis of the two-level correlations and the eigenfunctions statistics we show that the above extension of the Rosenzweig-Porter model indeed contains not one but two transitions. One of them  at $\gamma=2$
corresponds to the transition from the extended to the localized states. However, the extended states emerging at $\gamma<2$ are not
ergodic: their support set contains infinitely many $N^{D_{1}}$ sites in the $N\to \infty$ limit, which, however, is a zero fraction of all sites,
since $D_{1}<1$. Such non-ergodic extended states on RRG are recently discussed in Ref.~\cite{Our-BL}. With further decrease of $\gamma$ the second
transition at $\gamma=1$ happens which is a transition from the non-ergodic extended states
to the ergodic extended states with $D_{1}=1$ similar to the
eigenstates of the GOE.

We prove this statement in three steps. As the first step we use the perturbative arguments to compute the statistics of wave function amplitude $|\psi(r_{o})|^{2}$ in a certain observation point $r_{o}$. We obtain a drastic change of the character of this distribution at $\gamma=1$ and $\gamma=2$ which is summarized in Fig.~\ref{Fig:falpha}. This result is fully confirmed by a numerical diagonalization of the Hamiltonian (see Figs.~\ref{Fig:falpha300}~--~\ref{Fig:falpha075}). It is also confirmed by the numerical analysis of the moments of random wave functions which determine their Shannon entropy and the support set dimension $D_{1}$ (see Fig.~\ref{Fig:moments}). Then we compute numerically the overlap of amplitudes for two different wave functions with the energy difference $\omega$ and find the scaling with $N$ of the Thouless energy $E_{Th}\sim N^{-z}$ which exponent $z$ changes abruptly at $\gamma=1$ and $\gamma=2$. Finally, we perform a rigorous calculation of the spectral form-factor
 which also shows the transition at $\gamma=1$ and $\gamma=2$ (see Fig.~\ref{Fig:R(u)}).
In the last section we compare the corresponding results for our model and for the RRG and demonstrate their similarity.
It allows us to unify both models in a special universality class of random hierarchical models which differs from the one realized in localization transition points of two- and three-dimensional Anderson models.
Further details concerning this model can be found in Supplementary Materials.

\section{Statistics of eigenfunction amplitudes}
As the off-diagonal matrix elements in Eq.~(\ref{model}) are small, one can employ the perturbation theory for computing the distribution function of the amplitudes $x=N|\psi(r_{o})|^{2}$. The first order perturbation theory gives:
\begin{gather}\label{first-order}
|\psi_{n}(r_{m})|^{2}= {|H_{nm}|^{2}}/{(H_{nn}-H_{mm})^{2}},\;\;\;\;\;(n\neq m)
\end{gather}
where the maximum of $\psi_{n}(r)$ is supposed to be at $r=r_{n}$.

The perturbative series converge absolutely if the typical off-diagonal matrix element $|(H_{nm})_{\rm typ}|\sim \lambda N^{-\gamma/2}$ times the coordination number $N$ is much smaller than the typical difference of the diagonal matrix elements $|(H_{nn}-H_{mm})_{\rm typ}|\sim W\gg \delta(N)$. Thus it converges absolutely for $\gamma>2$ irrespectively of the statistics of diagonal matrix elements. For $\gamma\leq 2$ the convergence of the series occurs only because of the random and independently fluctuating signs of $H_{nm}$ and $(H_{nn}-H_{mm})$. Although it is hard to prove such a convergence rigorously, a plausible argument in its favor is that the effective coordination number of oscillatory contributions is $\sqrt{N}$ rather than $N$. The corresponding criterion of convergence is $\lambda\,N^{-\gamma/2}\,\sqrt{N}\ll W$ which is satisfied at $\gamma>1$.

Consider the regular part of the characteristic function $Q(\xi)= \langle e^{i \xi\,N|\psi(r_{o})|^{2}}\rangle$.
For the Gaussian distribution of matrix elements Eq.~(\ref{model}) we obtain $Q(\xi)= \mathcal{Q}(\xi N \sigma)$, where:
\begin{gather}\label{reg}
\mathcal{Q}(\zeta) = e^{-i\zeta/2}\, {\rm Erfc}\left(\sqrt{{-i\zeta}/{2}}\right)\approx 1-\sqrt{{-2 i \zeta}/{\pi}}.
\end{gather}
The function $P(x)=\int_{-\infty}^{\infty}e^{-i\zeta\,\frac{x}{N\sigma}}\,\mathcal{Q}(\zeta)\,\frac{d\zeta}{2\pi N \sigma}$ at $x\gg N\sigma\sim O(N^{1-\gamma})$ is dominated at $\gamma>1$ by small $\zeta\ll 1$. That is why it is only the expansion of Eq.~(\ref{reg}) at small $\zeta$ what matters for $P(x)$ at $\gamma>1$. Thus we obtain for the regular part of the eigenfunction distribution $P(x)$:
\begin{gather}\label{reg-P}
\mathcal{P}(x)=(\sqrt{2}\pi)^{-1}\,{(N\sigma)^{1/2}}/{x^{3/2}}.
\end{gather}
There are two normalization conditions for $P(x)$: the normalization of probability Eq.~(\ref{P-norm}) and the normalization of the wave function Eq.~(\ref{psi-norm}):
\bea\label{P-norm}
& &\int_{0}^{\infty}P(x)\,dx=1,\\
& &\int_{0}^{\infty}x \,P(x)\,dx=1\label{psi-norm}.
\eea
Eq.~(\ref{P-norm}) imposes a cut-off $x_{\min}\sim N^{-(\gamma-1)}$ to Eq.~(\ref{reg-P}) at small $x$, while Eq.~(\ref{psi-norm}) determines the upper cut-off $x_{\max}\sim N^{\gamma-1}$. A caution, however, should be taken: by normalization $\sum_{i}|\psi(r_{i})|^{2}=1$ the amplitude $|\psi(r_{i})|^{2}\leq 1$ on any lattice site
cannot exceed 1, and therefore $x\leq N$. One can see that the above estimation for $x_{\max}$ is valid only for $\gamma< 2$ when $N^{\gamma-1}\ll N$. For
$\gamma>2$ a correct $x_{\max}=N$. In order to compensate for the deficiency of normalization in Eq.~(\ref{psi-norm}) one has to assume a {\it singular} part of
$P(x)=\mathcal{P}(x)+A\,\delta(x-N)$. One can see that for $\gamma>2$ Eq.~(\ref{psi-norm}) is dominated by the singular term, and $A=N^{-1}$. This corresponds to the strongly localized wave functions. The mechanism of emergence of the singular term at the AL transition at $\gamma=2$ is somewhat similar to the Bose-condensation, where the singular term also appears because of the deficiency of normalization of the Bose-Einstein distribution.

\begin{figure}[t]
\center{\includegraphics[width=0.9\linewidth]{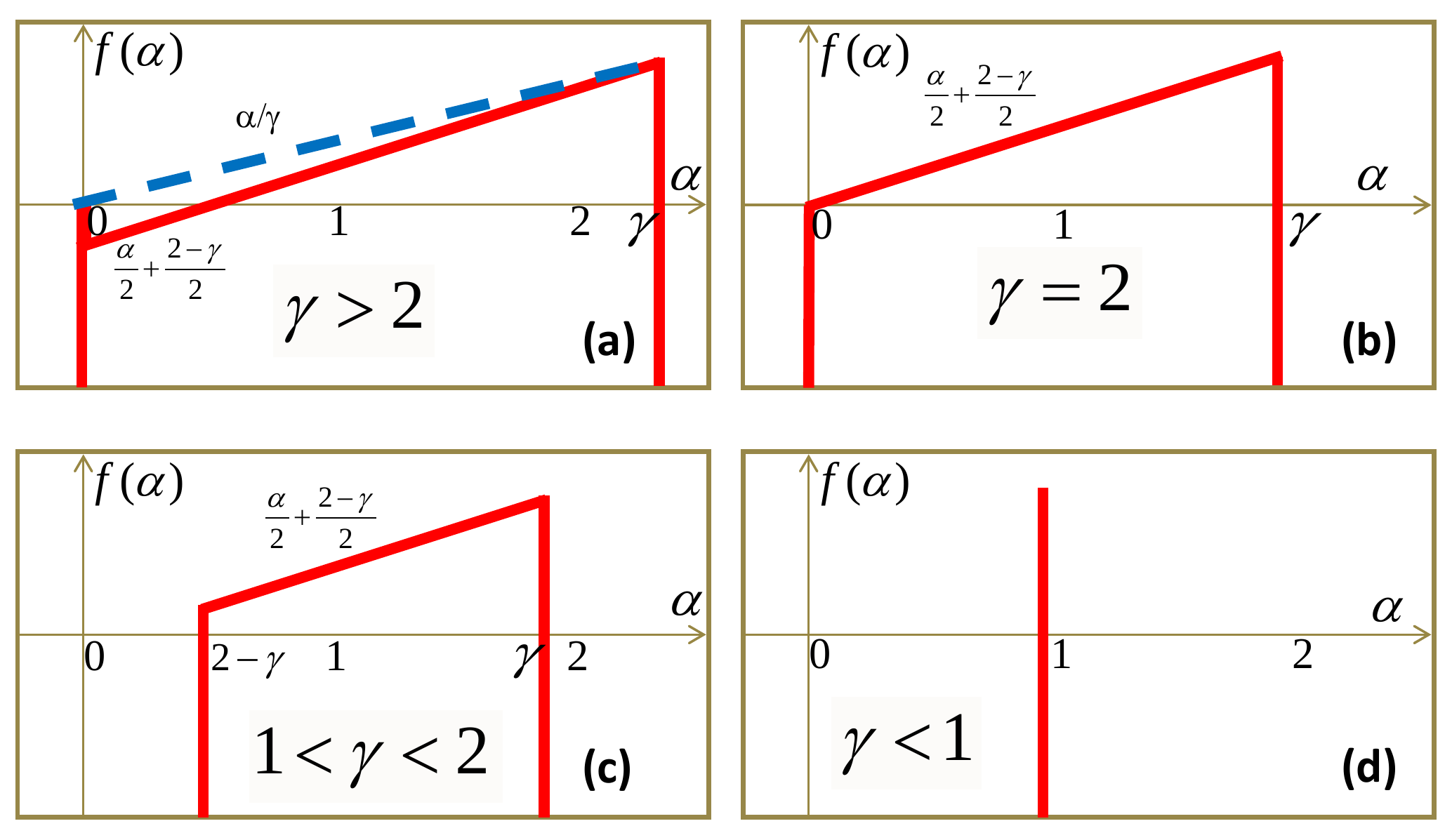}}
\caption{(Color online) The spectrum of fractal dimensions:
(a) the singular spectrum in the localized phase at $\gamma >2$. It corresponds to the same exponent $\tau_{q}$ as for $f(\alpha)$ shown by the dashed line.
(b) the triangular spectrum at the localization transition point $\gamma=2$.
(c) the spectrum with the gap $\alpha_{\min}=2-\gamma$ for the intermediate phase $1<\gamma<2$;
(d) the ergodic transition at $\gamma=1$ corresponds to the collapse of $\alpha_{\max}-\alpha_{\min}=2(1-\gamma)$.
}
 \label{Fig:falpha}
\end{figure}

One can express the distribution function Eq.~(\ref{reg-P}) through the {\it spectrum of fractal dimensions} \cite{MirRev,Our-BL}:
\begin{multline}\label{f-alpha-N}
f(\alpha)=\lim_{N\to \infty}f(\alpha,N)=\lim_{N\to \infty}\ln[x N \mathcal{P}(x)]/\ln N,
\end{multline}
where $\alpha=1-\ln x/\ln N$ or $|\psi(r_{o})|^{2}=N^{-\alpha}$. Using Eqs.~(\ref{reg-P})-(\ref{f-alpha-N}) one obtains:
\begin{gather}\label{falpha}
f(\alpha)={\alpha}/{2}+1-{\gamma}/{2},\;\;(\alpha_{\min}<\alpha <\alpha_{\max}).
\end{gather}
The upper cutoff $\alpha_{\max}=\gamma$ corresponds to the lower cutoff $x_{\min}$. The lower cutoff $\alpha_{\min}$ depends on $\gamma$. In the localized region $\gamma>2$, Fig.~\ref{Fig:falpha}(a), $\alpha_{\min}=0$. At the AL transition point $\gamma=2$ the function $f(\alpha)$ has the same triangular shape as at $W=W_{AT}$ on RRG, Fig.~\ref{Fig:falpha}(b). In the region of the extended non-ergodic states $1<\gamma<2$, Fig.~\ref{Fig:falpha}(c), $\alpha_{\min}=2-\gamma>0$. It is remarkable that
in the entire region $1 \leq\gamma\leq 2$ the symmetry \cite{MirFyod,MirRev} $f(1+x)=f(1-x)+x$ holds. Finally, at $\gamma=1$ the two limits $\alpha_{\min}$ and $\alpha_{\max}$ collapse in one point $\alpha=1$ which marks the transition point to the ergodic state, Fig.~\ref{Fig:falpha}(d).

Note that $f(\alpha)$ for $\gamma>2$ (see Fig.~\ref{Fig:falpha}(a)) has a singular peak at $\alpha=0$ which corresponds to the singular term $N^{-1}\delta(x-N)$ in $P(x)$.  This singular $f(\alpha)$ is not a limit of any {\it convex} function.
However, one may easily see that all the moments $N\langle |\psi|^{2q} \rangle \sim N^{-\tau_{q}}$ have the same exponents $\tau_{q}$ as for the ``convex'' $f(\alpha)$ shown by the dashed line in Fig.~\ref{Fig:falpha} (a): $f(\alpha)=\alpha/\gamma$ for $0<\alpha<\gamma$. Such a triangular $f(\alpha)$ with the slope smaller than 1/2 also holds in the localized phase on RRG \cite{Our-BL}.

\begin{figure}
\center{\includegraphics[width=0.9\linewidth]{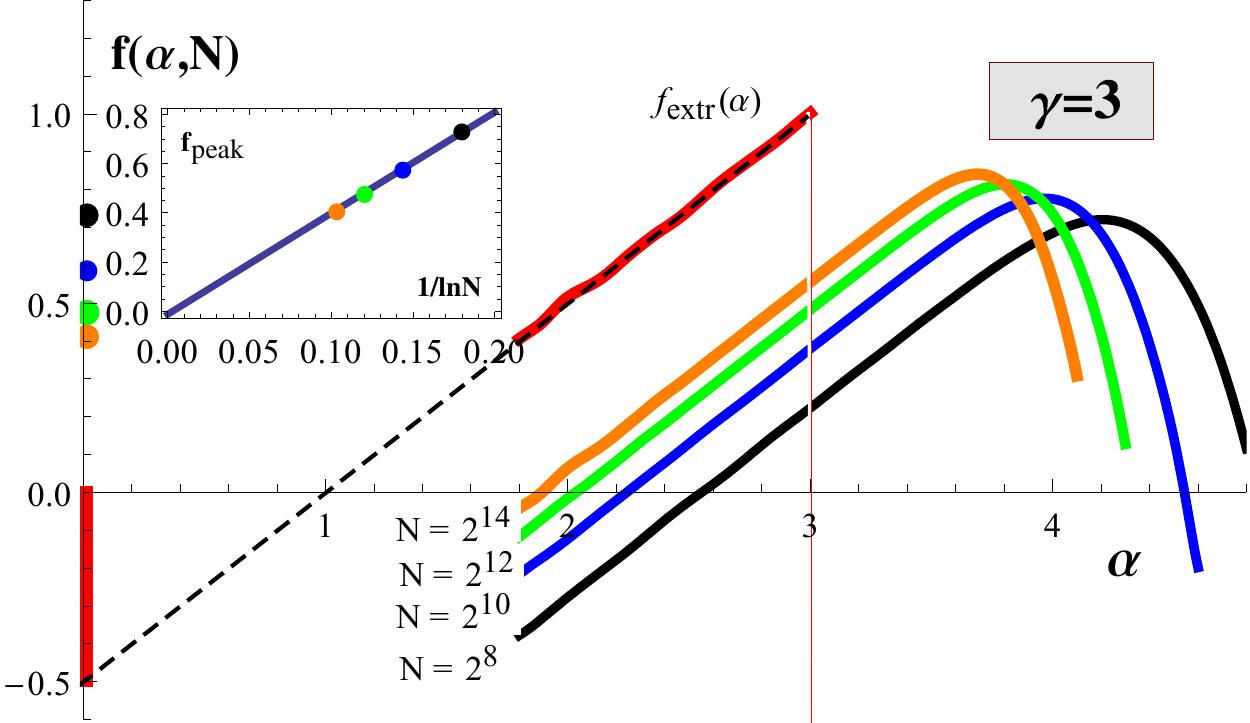}}
\caption{(Color online) Spectrum of fractal dimension for $\gamma=3$ obtained  numerically as in Ref.~\cite{Our-BL}. The linear part of the extrapolated $f(\alpha)$ (solid red line) is exactly as expected $f(\alpha)$ (black dashed line).
 The curves for $f(\alpha,N)$ for increasing $N$ are shown by black, blue, green and orange lines from bottom to top. The top of the singular peak at $\alpha=0$ shown by the points of the corresponding color, extrapolates to zero as expected (see inset);
(inset) the $1/\ln N$ extrapolation of the singular peak value $f_{peak}=f(0,N)$.
   }
\label{Fig:falpha300}
\end{figure}

\begin{figure}
 \center{\includegraphics[width=0.9\linewidth]{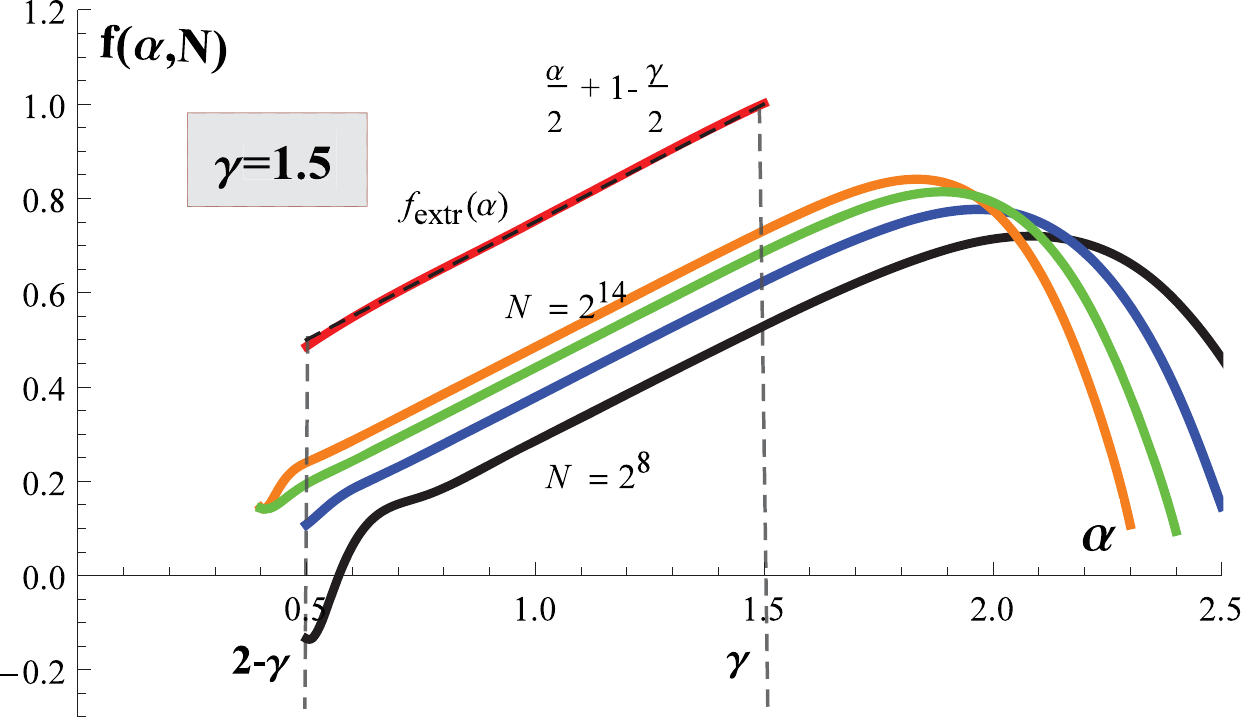}}
 \caption{(Color online) Spectrum of fractal dimension in the intermediate phase for $\gamma=1.5$ obtained numerically as in Ref.~\cite{Our-BL}. All notations are the same as in Fig.~\ref{Fig:falpha300}.  Line colors correspond to the same values of $N$ as in Fig.~\ref{Fig:falpha300}. Expected $f(\alpha)$ is shown by a black dashed line.
 } \label{Fig:falpha150}
 \end{figure}

\begin{figure}
 \center{\includegraphics[width=0.9\linewidth]{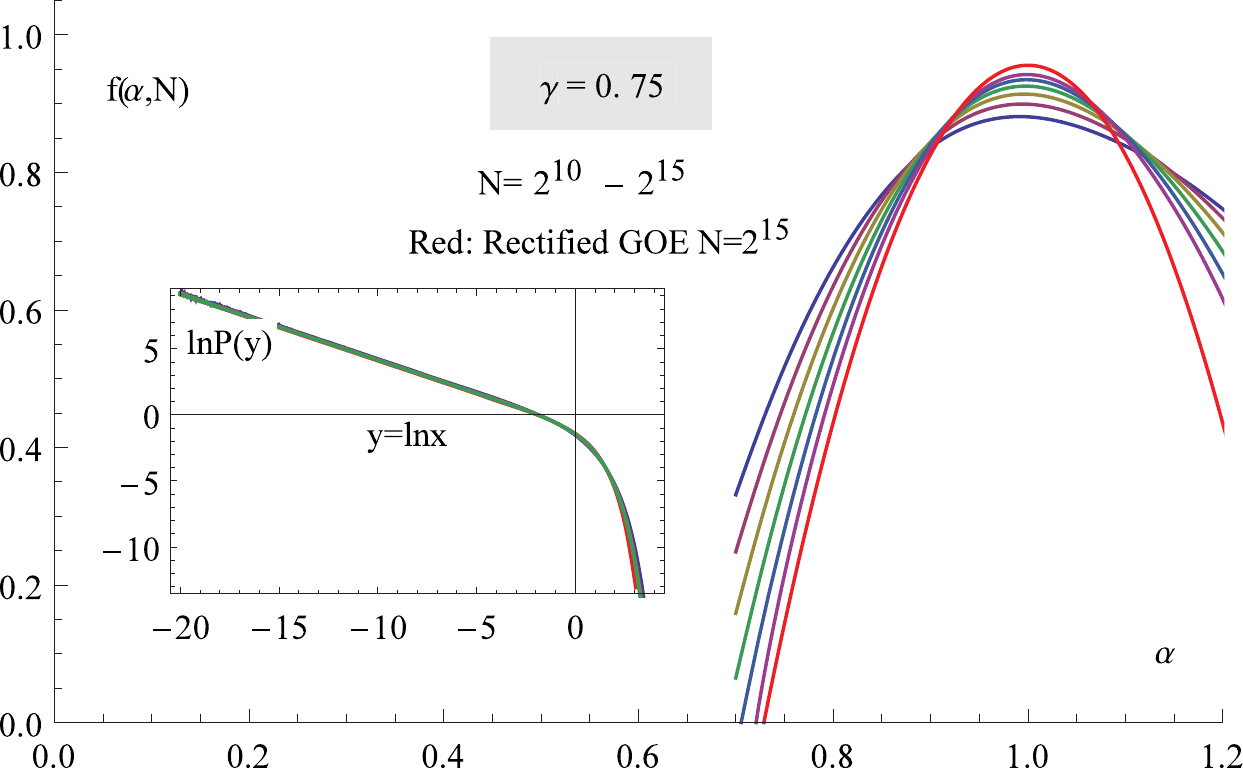}}
 \caption{(Color online) Finite-N spectrum of fractal dimensions $f(\alpha,N)$ for $\gamma=0.75$ and $N=2^{10}-2^{15}$ obtained by the rectification procedure of Ref.~\cite{Our-BL}. For comparison we also present $f(\alpha,N)$ (shown by a red line) for the Porter-Thomas distribution of wave function amplitudes in the GOE obtained by the same procedure at $N=2^{15}$. It almost coincides with the (violette) curve for $f(\alpha,N)$ computed at the same $N=2^{15}$ for our model with $\gamma=0.75$. In the inset: $\ln P(x)$ vs $\ln x$ for the same $\gamma=0.75$ and system sizes as in the main plot. The corresponding curve for GOE is shown in red. All the curves are almost indistinguishable.
 } \label{Fig:falpha075}
 \end{figure}

\begin{figure}[t]
\center{
\includegraphics[width=0.9\linewidth]{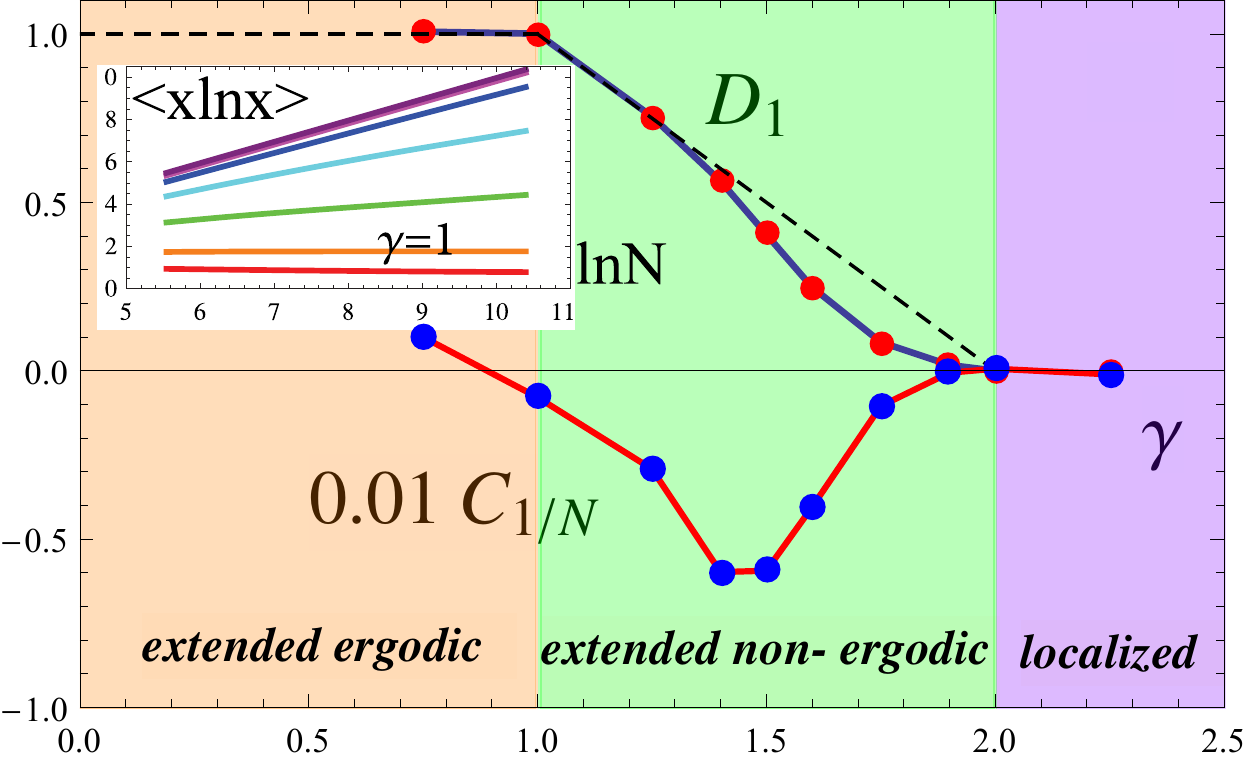}
}
\caption{(Color online) The support set dimension $D_{1}(\gamma)$ and the global curvature $C_{1/N}(\gamma)$ extracted from the fit $\langle x \ln x\rangle = (1-D_{1})\,\ln N + C_{0} + C_{1/N}\,N^{-1}$ vs. $\gamma$. The dashed line is the prediction for $D_{1}$, Eq.~(\ref{fr-dim}); (inset) The average $\langle x \ln x \rangle$ vs $\ln N$ for $\gamma$ from $0.75$ (bottom) to $2.25$ (top) with steps $0.25$. It is related with the Shannon entropy $-\sum_{r}|\psi(r)|^{2}\,\ln |\psi(r)|^{2}=\ln N -\langle x \ln x\rangle$.
The global curvature $C_{1/N}$ changes sign at the transition points $\gamma=1$ and $\gamma=2$.
}
 \label{Fig:moments}
\end{figure}

The numerical calculation of $f(\alpha)$ (see Sec.~\ref{SMsec:f(a)} in the Supplementary Materials) which involve the rectification and extrapolation procedures described in Ref.~\cite{Our-BL}, fully confirms the above results. In Fig.~\ref{Fig:falpha300} we present the results of this calculation for $N=10^{8}-10^{14}$ and  the extrapolated $f(\alpha)$ (shown by a solid red line) for $\gamma=3$ which perfectly coincides with the prediction of our perturbative analysis above. The similar coincidence was obtained for $\gamma=1.5$, while for $\gamma=0.75$ the distribution function
$P(x)$ is practically indistinguishable from the Porter-Thomas distribution of the GOE.

 \section{ The support set dimension $D_{1}$}
By calculating the Legendre transform  $\tau_{q}$ \cite{MirRev}  of $f(\alpha)$ (shown in Fig.~\ref{Fig:falpha}) one finds that in the intermediate phase $1<\gamma<2$ all fractal dimensions $D_{q}=\tau_{q}/(q-1)$ for $q>1/2$ are the same and equal to (see Sec.~\ref{SMsec:moments} in Supplementary Materials for details):
\be\label{fr-dim}
D_{q}=D=2-\gamma,\;\;\;\;(q>1/2).
\ee
Thus the support set of a typical wave function is a fractal containing $ N^{D_{1}}=N^{2-\gamma} $ sites. As $N^{2-\gamma}\to\infty$ in the limit $N\to\infty$ it is an extended state. However the support set contains a fraction of all sites $F=N^{1-\gamma}$ tending to zero in this limit. Thus it is a non-ergodic state.

In the localized phase $\gamma>2$ (including the critical point $\gamma=2$)  we obtained:
\begin{eqnarray}\label{tau-freez}
\tau(q)=\left\{\begin{array}{lr}\gamma q-1, & q<1/\gamma \cr
0, & q>1/\gamma \cr\end{array}\right.
\end{eqnarray}
 One can see that the fractal dimensions $D_{q}=0$ only at $q>1/\gamma$, while they are non-zero and negative for $0<q<1/\gamma$.
This is not exactly the behavior of the typical Anderson insulator where {\it all} fractal dimensions with $q>0$ are equal to zero. The behavior similar to Eq.~(\ref{tau-freez}) were found in certain two-dimensional random Dirac models \cite{frozen1,frozen2,frozen3,Foster}. Such a quasi-localized phase is referred to as the {\it frozen} phase and the corresponding transition is known as the {\it freezing transition}. In such a phase a typical wave function amplitude has several sharp peaks separated by valleys where $|\psi|^{2}$ is not exponentially but only power-law small in $N$ ($|\psi|_{\rm typ}^{2}>N^{-\gamma}$ in our case). The same behavior is also found for the RRG \cite{Our-BL} with $W>W_{AT}$.

In order to check the existence of the intermediate phase numerically we computed the average $\langle x \ln x \rangle$ which is directly related with the Shannon entropy and the dimension $D_{1}$ of the support set of fractal wave functions \cite{KravAltScard}. The results are shown in the inset of Fig.~\ref{Fig:moments} where $N$ span from $256$ up to $32768$. The corresponding values of $D_{1}$ extracted from the linear in $\ln N $ fit are shown in Fig.~\ref{Fig:moments} which are consistent with the transitions at $\gamma=2$ and $\gamma=1$. The deviation from the expected $D_{1}=2-\gamma$   shown by a dashed line in Fig.~\ref{Fig:moments} is a finite-size effect.
Indeed, the correlation volume $N_{c}$ close to the localization transition at $\gamma=2$ is exponentially large $N_{c}\propto e^{c/|2-\gamma|}$. This follows from Eq.~(\ref{S-u}) of Sec.~\ref{Sec5} where the Poisson limit is reached only for $N^{\gamma-2}\gg O(1)$ or $\ln N \gg \ln N_{c} \sim 1/(\gamma-2)$ (see Sec.~\ref{SMsec:support_set} of Supplementary Materials for details). The similar exponential dependence $N_{c}\sim e^{c/\sqrt{|W-W_{c}|}}$ of the correlation volume was obtained on the Bethe lattice \cite{Fyod-corr}. For system sizes $N\ll N_{c}$ one should see the properties of the critical point $\gamma=2$ where $D_{1}=0$. Thus in the vicinity of the transition point the support set dimension  extracted from the finite-size simulations should show a tendency towards smaller values as in Fig.~\ref{Fig:moments}. However, for $\gamma<1.5$ at our systems sizes the support set dimension $D_{1}$ approaches the values expected in $N\to \infty$ limit (dashed line in Fig.~\ref{Fig:moments}). This fact implies that for $\gamma<1.5$ we reached $N\gg N_{c}$ and thus it may serve as numerical evidence of convergence and existence of non-ergodic extended phase in the thermodynamic limit.

We also introduced the $1/N$ corrections to the fit with its magnitude $C_{1/N}$ being a measure of the global curvature of the $\langle x \ln x \rangle$ vs. $\ln N$ dependence (see Sec.~\ref{SMsec:support_set} of Supplementary Materials for details). Remarkably, $C_{1/N}$ changes sign at both the transition points $\gamma=1$ and $\gamma=2$ (though the positive $C_{1/N}$ is very small for $\gamma>2$).
We also checked that it changes sign at the localization transition point of the 3D Anderson model (not shown).
We believe that the changing of sign of $C_{1/N}$ is a convenient way to identify the points of both localization and ergodic transitions.
\begin{figure}
\center{
\includegraphics[width=0.9\linewidth]{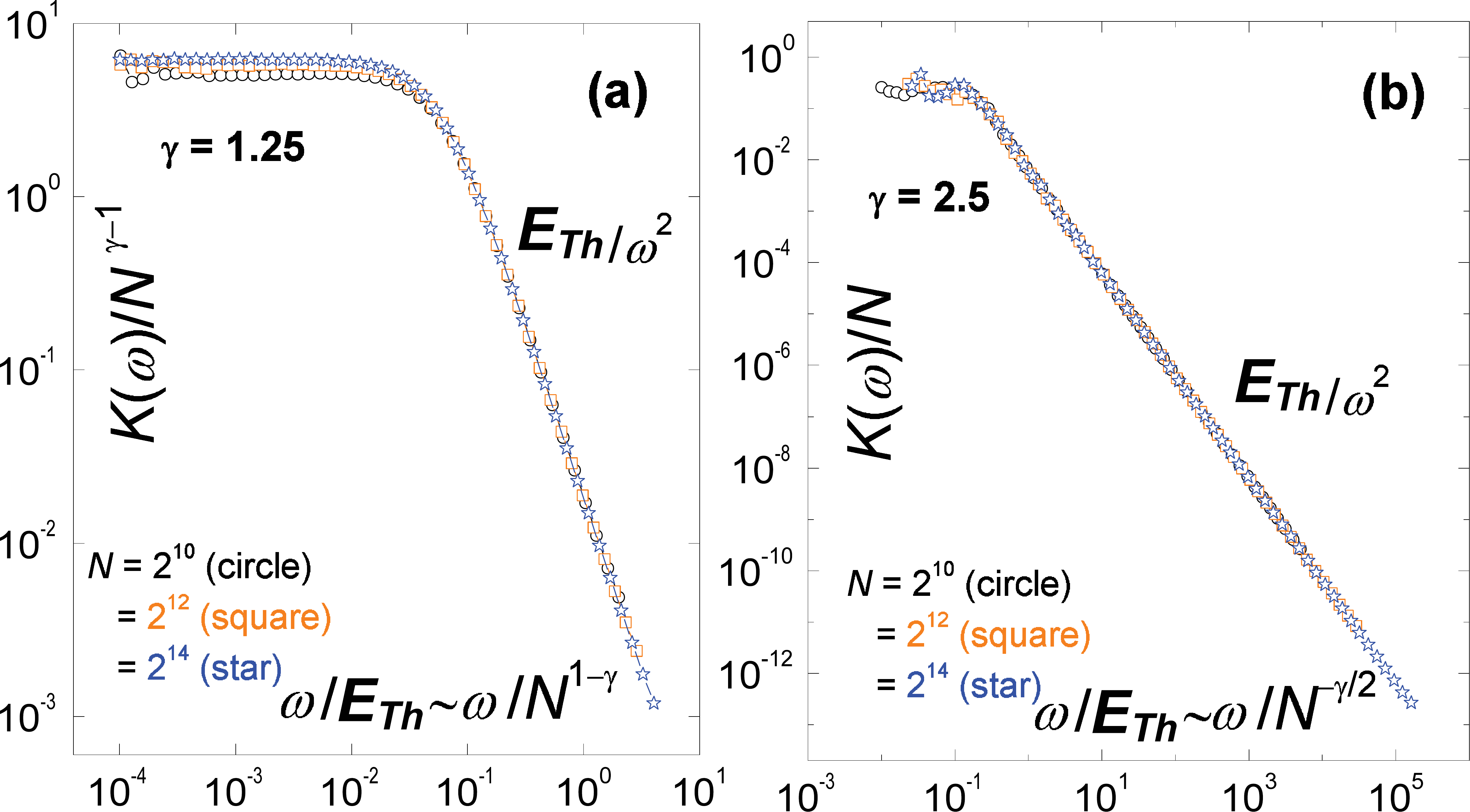}
}
\caption{(Color online) The overlap correlation function $K(\omega)=N\sum_{r} \langle |\psi_{E}(r)|^{2}\,|\psi_{E+\omega}(r)|^{2}\rangle $.
(a) For $1<\gamma<2$ functions $K(\omega)N^{1-\gamma}$ for different $N$ collapse into the same curve in coordinates $\omega/E_{Th}\propto \omega/N^{1-\gamma}$ (see also an inset to Fig.~\ref{Fig:R(u)}). (b) For $\gamma\geq 2$ the collapse of $K(\omega)/N$ occurs in coordinates
$\omega/N^{-\gamma/2}$.
}
 \label{Fig:overlap}
\end{figure}

\section{Overlap of different wave functions}
Next we compute numerically the overlap of different wave functions $K(\omega)=N\sum_{r}\langle |\psi_{E}(r)|^{2}|\psi_{E+\omega}(r)|^{2}\rangle$.
Note that for the ergodic wave functions of GOE $K(\omega)=1$ is independent of the energy difference $\omega$, as in this case the overlap is always $100\%$ (see Sec.~\ref{SMsec:overlap} of Supplementary Materials for details).
Our results presented in Fig.~\ref{Fig:overlap} show that for $\gamma>1$ the overlap $K(\omega)$ has a plateau at $\omega<E_{Th}$ which is followed by a fast decrease $K(\omega)\propto 1/\omega^{2}$ for $\omega>E_{Th}$. The Thouless energy $E_{Th}$ \cite{Altshuler_Shklovskii} that separates the GOE-like behavior (plateau) from the system specific behavior ($K(\omega)\propto \omega^{-2}$), depends on $N$ as a power-law $N^{-z}$. However, the scaling exponent $z$ is different in all the three phases (see Fig.~\ref{Fig:overlap}). In the localized phase Fig.~\ref{Fig:overlap}(b) we obtained $z=\gamma/2$ which corresponds to rare resonances when $\omega<|H_{n\neq m}|\sim N^{-\gamma/2}$. In the extended non-ergodic phase Fig.~\ref{Fig:overlap}(a) we found $E_{Th}\sim N^{1-\gamma}\sim \delta\,N^{D}$, where $\delta = 1/(N\,p(0))$ is the mean level spacing.
This corresponds to all $N^{D}$ sites in the support set being in resonance with each other.
The corresponding $N^{D}$ states produced by  linear combinations of basis states localized on resonant sites form a mini-band of levels of the width $E_{Th}\sim \delta\,N^{D}$.
Clearly, the states inside such a mini-band should have the GOE-like correlations.
On the contrary, the states separated by the energy distance $\omega>E_{Th}$ should belong to different support sets which poorly overlap with each other.
At the ergodic transition at $\gamma=1$ and in the entire extended ergodic state at $\gamma<1$ we obtain $E_{Th}\sim O(N^{0})$ (see Sec.~\ref{SMsec:overlap} of Supplementary Materials for details), and the plateau extends to entire spectral band-width.
The emergence of such a plateau that survives the limit $N\to\infty$ is a signature of the ergodic state \cite{CueKrav-orig}.

Surprisingly, the overlap function $K(\omega)\sim N^{1-\gamma}/\omega^{2}\to 0$ as $N\to\infty$ at any fixed $\omega$ and $\gamma>1$. This phenomenon of ``repulsion of wave functions'' \cite{CueKrav-orig} is a peculiar feature of our model.
The non-ergodic fractal states at the localization transition points of the two and three- dimensional Anderson models of the Dyson symmetry classes, as well as those of the power-law banded random matrices \cite{Seligman, MirRev, CueKrav-orig} show a different behavior.
In these models, the Thouless energy for fractal states is proportional to the mean level spacing $E_{Th}\sim \delta$ and the behavior for $\omega>E_{Th}(N)$ is described by the conventional Chalker's scaling  \cite{Chalk-Daniel,KrMut,KrOsYev}:
\be\label{Chalker}
K(\omega)\sim \omega^{-1+D_{2}}.
\ee
Only at a very large energy separations $\omega$ of the order of the total spectral band-width, the ``repulsion of wave functions'' was observed \cite{CueKrav-orig}.

A remarkable feature of the present model is that the Thouless energy in the region of extended non-ergodic states is much larger than the mean level spacing:
\be\label{large-ETh}
E_{Th}\sim \delta\,N^{D_{2}}\propto N^{-z}.
\ee
One can interpret this relationship as a non-trivial dynamical scaling exponent
\be
z=1-D_{2}<1.
\ee
For non-interacting systems in two or three-dimensions in the point of Anderson transition $z=1$ for all Dyson symmetry classes.
A non-trivial $z$ is known only in two-dimensional systems described by the Dirac equation with random vector-potential which belong to chiral symmetry classes \cite{frozen1,frozen2,frozen3,Foster} where the freezing transition is observed.

In terms of the dynamical exponent $z$ the  leading power-law  term in the Chalker's scaling for $\omega\gg E_{Th}$ can be rewritten as \cite{Foster}:
\be\label{Chalker-z}
K(\omega)\propto \omega^{-\mu},\;\;\;\;\;\mu=(1-D_{2})/z.
\ee
In our model we have:
\be\label{fall}
K(\omega)\sim \frac{E_{Th}}{\omega^{2}},\;\;\;\;(\omega\gg E_{Th}).
\ee
One can consider Eq.~(\ref{fall}) as a particular case of expansion in $E_{Th}/\omega\ll 1$ with the leading term exponent $\mu=1$:
\be\label{fall-gen}
K(\omega)=\frac{1}{\omega}\,\left[ c_{0}+c_{1}\,\frac{E_{Th}}{\omega}+...~\right],
\ee
in which the coefficient $c_{0}$ is zero. We will see below that Eqs.~(\ref{large-ETh}, \ref{fall-gen}) hold for the RRG too. However in this case the coefficient $c_{0}\sim 1$ is non-zero. Thus one can speak of the special universality class of models with a non-trivial $z\ne 1$ and $\mu=1$ which the present model belongs to together with the RRG model.

\section{Spectral form-factor}\label{Sec5}
Finally, we present the results of a rigorous calculation of the spectral form-factor $C(t,t')=\sum_{n\neq m}e^{it E_{m}+it' E_{n}}$, with a set of eigenvalues $\{E_{n}\}$ of $H$. To this end we generalize to the case $\gamma>1$  the results of Ref.~\cite{ShapiroKunz} where $C(t,t')$ was derived for URP model Eq.~(\ref{model}) with $\gamma=2$ using the Itzykson~-~Zuber formula of integration over unitary group.  The final result (see details of the derivation in Sec.~\ref{SMsec:form-factor} of Supplementary Materials) for $C(t,t')=2\pi \delta(t+t')\,[S(\frac{E_{Th}}{2}(t-t'))-1]$ is given by Eq.~(\ref{S-u})
\begin{widetext}
\be\label{S-u}
S(u)=
 1 + e^{-2 \pi \Lambda^{2} u}e^{-\Lambda^{2} u^{2} N^{\gamma - 2}}\left[ \frac{
 2 I_{1}( \kappa u^{3/2}) }{\kappa u^{3/2}}
 -
 \frac{1}{4 \pi} \kappa u^{5/2}\,N^{\gamma-2}\,
 \int_{0}^{\infty}\frac{x\,dx}{\sqrt{x+1}}\,I_{1}(\kappa u^{3/2}\sqrt{x+1})\,e^{-x\,u^{2}\Lambda^{2}N^{\gamma-2}}\right],
\ee
\end{widetext}

with the modified Bessel function $I_{1}(x)$, $E_{Th}=\delta\,N^{2-\gamma}$, $\kappa=\sqrt{8\pi N^{\gamma-2}}\Lambda^{2}$ and $\Lambda=\lambda p(0)$.
\begin{figure}[h]
\center{
\includegraphics[width=0.7\linewidth]{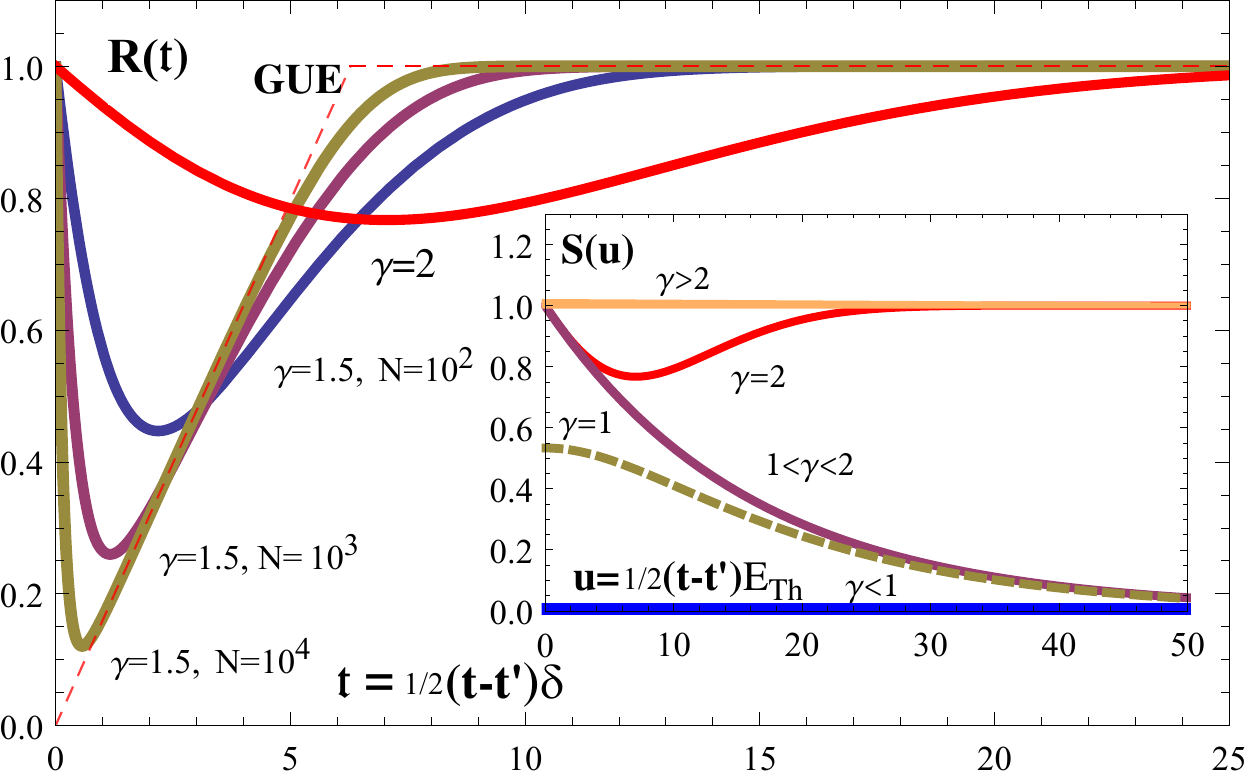}
}
\caption{(Color online) Unfolded spectral form-factor $R(\mathfrak{t})$
for the RP model for two different cases: (i) $\gamma=2$ and (ii) $\gamma=1.5$ at $N=10^{2}, 10^{3}, 10^{4}$. The falling part corresponds to attraction of levels while the rising part corresponds to repulsion of levels. The GUE form-factor is shown by the red dashed line. The Poisson distribution corresponds to $R(\mathfrak{t})=1$. (inset) The spectral form-factor in the variable $u=\frac{1}{2}(t-t')\,E_{Th}$ with $E_{Th}=\delta\,N^{2-\gamma}$ in the $N\to\infty$ limit for different values of $\gamma$. There are five distinctly different phases:  insulator $\gamma>2$, AT critical $\gamma=2$, non-ergodic extended $1<\gamma<2$, ET critical $\gamma=1$ and ergodic $\gamma<1$.
}
 \label{Fig:R(u)}
\end{figure}
The unfolded spectral form-factor $R(\mathfrak{t})$ with $\mathfrak{t}=\frac{1}{2}(t-t')\delta$ is given by $R(\mathfrak{t})=S(\mathfrak{t}N^{2-\gamma})$.
It follows from Eq.~(\ref{S-u}) that for $\gamma>2$, $R(\mathfrak{t})\to 1$  in the $N\to \infty$ limit, which corresponds to completely uncorrelated energy levels and the exact Poisson statistics.
Another important feature of Eq.~(\ref{S-u}) is that $R(0)=1$ for all $\gamma>1$.

In Fig.~\ref{Fig:R(u)} we plot the unfolded spectral form-factor $R(\mathfrak{t})$ for the two phases: (i) the critical phase of the AL transition at $\gamma=2$ and (ii) the intermediate phase at $1<\gamma<2$.  One can see that while at $\gamma=2$ the function $R(\mathfrak{t})$
has a non-trivial $N\to\infty$ limit, for $1<\gamma<2$ the limit coincides with that of the GUE, except for the point $\mathfrak{t}=0$ where there is a jump in $R(\mathfrak{t})$. This jump is a hallmark of the intermediate phase. To demonstrate this more clearly we blow up the region of small $\mathfrak{t}$ by re-scaling the variable $\mathfrak{t}\Rightarrow u=\frac{1}{2}(t-t')\delta\,N^{2-\gamma}=\frac{1}{2}(t-t')\,E_{Th}$. Note that $E_{Th}=\delta N^{2-\gamma}$ appears again as the characteristic scale where the level repulsion is taken over by the level attraction in Fig.~\ref{Fig:R(u)}.
In this new variable $R(u N^{\gamma-2})=S(u)$ has a non-GUE $N\to\infty$ limit:
\be\label{Lorenz}
S(u)=
\exp(-2\pi \Lambda^{2}\,u),\;\;\;\;1<\gamma<2,
\ee
which is shown in the inset to Fig.~\ref{Fig:R(u)}. The true GUE form factor is just identically zero in this limit.
The existence of the new scale $E_{Th}$ and a non-GUE $N\to \infty$ limit Eq.~(\ref{Lorenz}) in the variable $u=\frac{1}{2}(t-t')\,E_{Th}$ have been overlooked in Ref.~\cite{BrezHik}.

Eq.~(\ref{Lorenz}) holds for $u> E_{Th}\sim N^{1-\gamma}$, and $S(u)$ is saturated at $S(0)\approx  e^{-2\pi \Lambda^{2}\,N^{1-\gamma}}$ for $u< N^{1-\gamma}$, which corresponds to $|t-t'|$ smaller than the inverse total spectral band-width (see Sec.~\ref{SMsec:form-factor} of Supplementary Materials for details).
For $\gamma=1$ we have $N^{1-\gamma}=1$ . Thus the value $S(0)\sim e^{-2\pi\,\Lambda^{2}}$ at $\gamma=1$ is smaller than 1. So, in addition to the specific critical behavior of $S(u)$ at $\gamma=2$ (shown by the red curve in Fig.~\ref{Fig:R(u)}) one obtains yet another critical behavior of $S(u)$ at the ergodic transition $\gamma=1$ (shown by the dashed yellow line in Fig.~\ref{Fig:R(u)})
which is stable in the $N\to\infty$ limit and is characterized by $S(0)<1$.
For $\gamma<1$, $N^{1-\gamma}$ is increasing with N, making $S(0) =\max S(u)\to 0$ as $N\to \infty$. This is how the GUE limit $S(u)\equiv 0$ is reached.

Note that the fact that $E_{Th}\gg \delta$ affects the level number variance ${\rm var}(n)$ ($n$ and ${\rm var}(n)$ are the average number of levels and the level number variance in a certain spectral window, respectively) in which a new scale $E_{Th}/\delta \sim N^{2-\gamma}$ appears for $n$ at $1<\gamma<2$:
\be\label{LNV}
{\rm var}(n)=\left\{ \begin{array}{ll} \sim \ln n, & 1\ll n\ll N^{2-\gamma}\cr
n, & N^{2-\gamma}\ll n \ll N. \end{array}\right.
\ee
The level compressibility $\chi$ \cite{ChalkKravLer,KrMut} is ill-defined in our model because of the jump in the spectral form factor $R(\mathfrak{t})|_{N\to \infty}$ at $\mathfrak{t}=0$. Formally it can take any value from $\chi=0$ to $\chi=1$ depending on the parameter $n/N^{2-\gamma}$. This is in contrast to other random matrix models with  multifractal eigenstates, e.g. PLBRM  and Moshe-Neuberger-Shapiro models (see Ref.~\cite{KrMut} and references therein) where the level compressibility is well defined and takes a definite value $0<\chi<1$ in the $N\to\infty$ limit.

\section{Comparison with RRG model}
In the Introduction we mentioned a heuristic relation between the Anderson model on a hierarchical RRG and the RP model which has no apparent hierarchical structure. It is instructive now to compare the main results of this paper with the corresponding results for RRG.
\begin{figure}[t]
\center{
\includegraphics[width=0.9\linewidth]{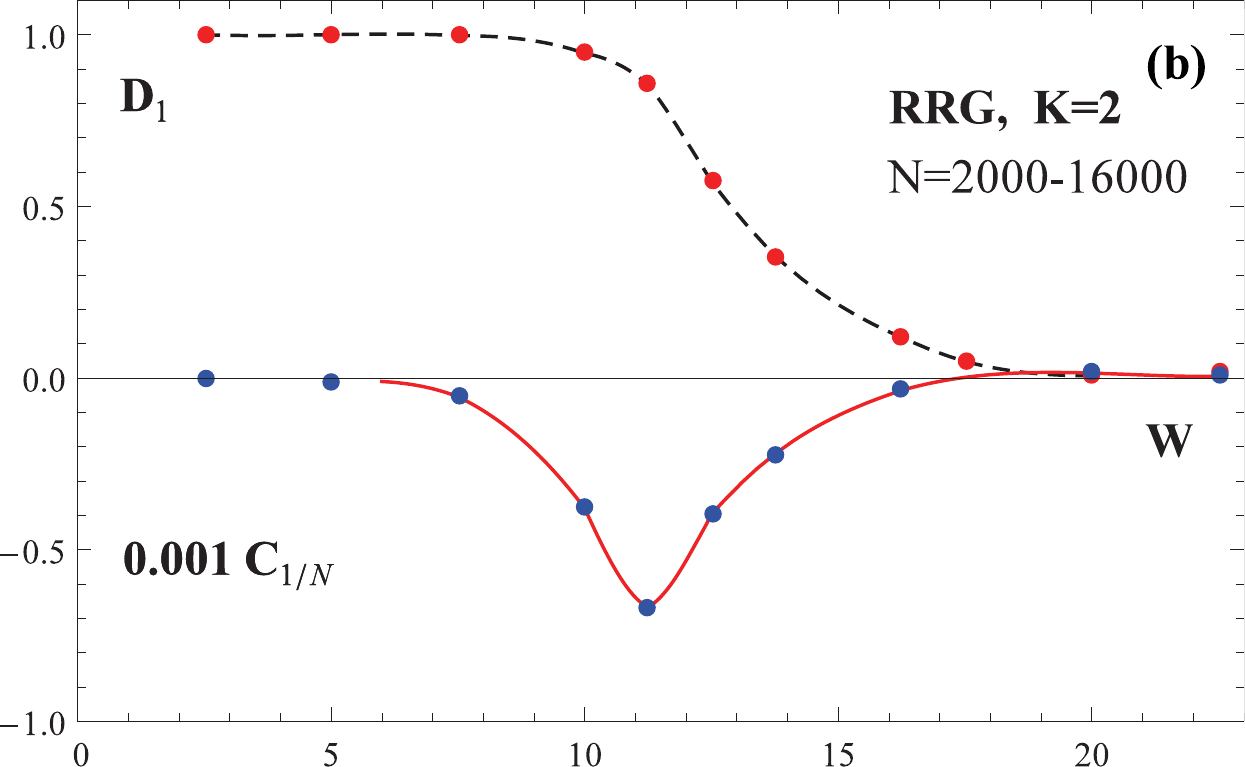}
}
\caption{(Color online) The support set dimension $D_{1}$  and the global curvature $C_{1/N}$ as the function of disorder strength $W$ extracted for the RRG model as in Fig.~\ref{Fig:moments}. No ergodic transition is detected by changing of sign of the global curvature. This is consistent with the statement \cite{Our-BL} that the entire extended phase is non-ergodic.
}
 \label{Fig:D1-C1N-RRG}
\end{figure}

\begin{figure}[h]
\center{
\includegraphics[width=0.9\linewidth]{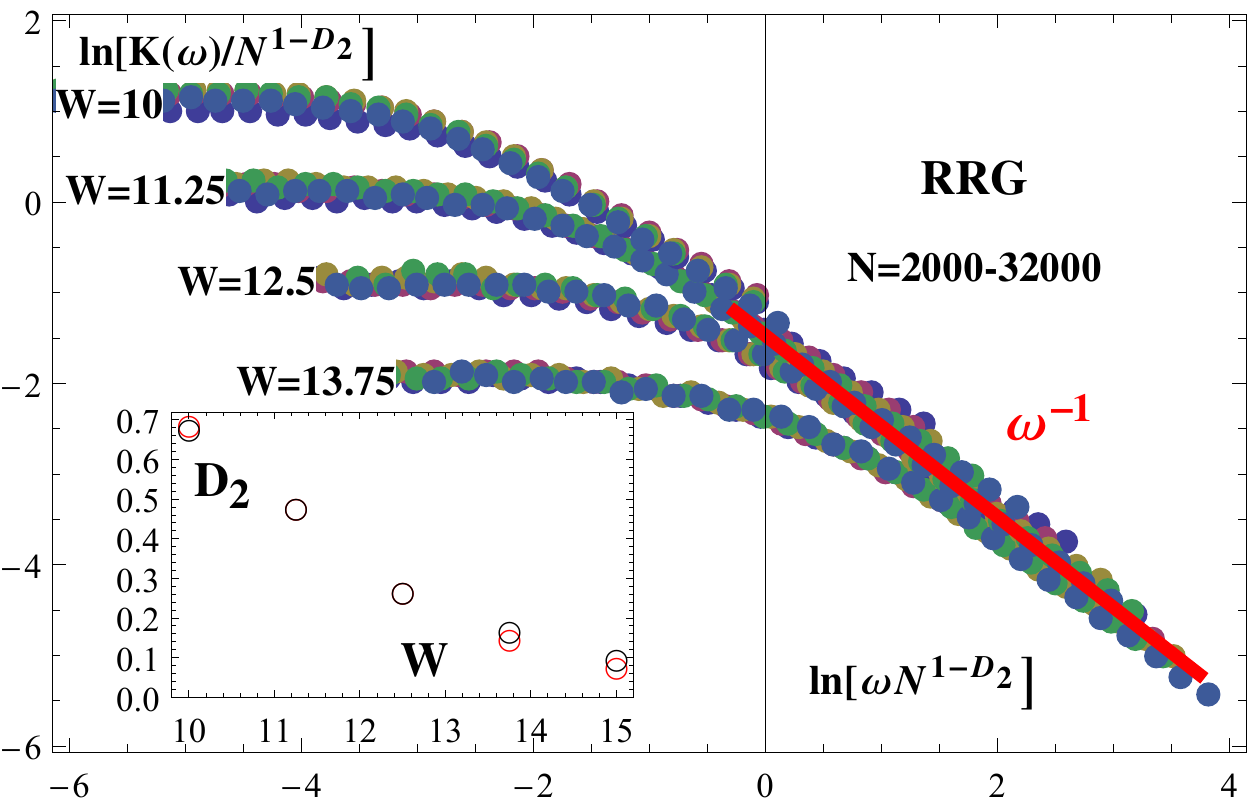}
}
\caption{(Color online) Collapse of data for $K(\omega)$ for the Anderson model on RRG with the branching number $K=2$ and $N=2000$, $4000$, $8000$, $16000$, $32000$. The collapse occurs in the coordinates $K(\omega)/N^{1-D_{2}}$ and $\omega/ N^{-1+D_{2}}$ which corresponds to Eqs.~(\ref{large-ETh}),(\ref{fr-dim}) and Fig.~\ref{Fig:overlap}(a) in RP-ensemble. The values of fractal dimension $D_{2}$ extracted from the best collapse are in excellent agreement with the corresponding values found from the moments $\langle |\psi|^{4} \rangle$ (see inset).
Different disorder strengths $W$ correspond to different collapse curves and different dynamical exponents $z$ in $E_{Th}=N^{-z}$. This excludes a possibility for an apparent non-ergodicity of extended phase to be a finite-size effect reflecting properties of only one single critical disorder strength. The behavior of $K(\omega)\sim 1/\omega$ for $\omega>E_{Th}=\delta\,N^{D_{2}}$ corresponds to Eq.~(\ref{fall-gen}) with a non-zero $c_{0}\sim 1$.
(inset) The fractal dimensions $D_{2}=1-z$ as a function of $W$ found from the best collapse of data for $K(\omega)$ (red circle) and from the scaling of moments $\langle |\psi|^{4} \rangle$ with $N$ (black circle).}
 \label{Fig:ETh-RRG}
\end{figure}
First of all we recall (see Fig.~\ref{Fig:falpha} and the corresponding explanations in the text) that all the moments  $\langle |\psi|^{2q}\rangle$ in the localized  and the AT critical phases of our model have exactly the same $q$-dependence  as in the corresponding phases of the RRG. The $N$- and $\gamma$- dependence of the moments is also very similar  (cf. Fig.~\ref{Fig:moments} with Fig.~\ref{Fig:D1-C1N-RRG}) to the corresponding $N$ and $W$-dependences for  the random wave functions obtained by the exact diagonalization of the Anderson model on RRG with the branching number $K=2$ and $N=2000-16000$. However, there is an important difference: we found only one point of changing the sign of $C_{1/N}$ on RRG which corresponds to the known point of the Anderson localization transition at $W\approx 17.5$.

In Fig.~\ref{Fig:ETh-RRG} we demonstrate that in the case of RRG the scaling of the Thouless energy with the system size follows the same Eq.~(\ref{large-ETh}) as for the present RP model and is different from the standard Chalker's scaling. The falling part of $K(\omega)$ at $\omega\gg E_{Th}$ can be described by the unified expansion Eq.~(\ref{fall-gen}) both for RRG and our model, albeit the coefficient $c_{0}$ is zero for the present model and is non-zero for RRG. It is important that for RRG the data for properly re-scaled $F(y)=N^{-z}K(y N^{-z})$ at different $N$
collapse on one same scaling curve $F(y)\equiv F_{W}(y)$  that, however,  depends on $W$, as well as the exponent $z=1-D_{2}$ (see inset of Fig.~\ref{Fig:ETh-RRG}). This is very different from the usual scaling  $K(\omega)=N^{a}\,F(\omega N^{z}, N_{c}/N)$ in the vicinity of a single critical point $W=W_{c}$ where the exponents $a$ and $z$ are determined by the property of this critical point and not by the distance $|W-W_{c}|$ from this point which determines only the correlation volume $N_{c}(|W-W_{c}|)$.    In our opinion, this implies that there is a {\it line of critical points}
at $W<W_{c}$ which determine the behavior of the system {\it at least} in a parametrically large interval of sizes $N_{c2}\gg N\gg N_{c}$, with the second characteristic size scale $N_{c2}\gg N_{c}$.

Our conclusion is that the localized and AT critical states are very similar for our model and RRG. The extended states show non-ergodicity in a broad interval of $\gamma$  and $W$ and are characterized by the Touless energy which in both models is much larger than the mean level spacing. However, the existence or non-existence of the ergodic transition is more subtle and depends on tiny features of the model.
It exists in our model and most probably does not exist on RRG with the branching number $K=2$.
Nonetheless our study largely confirms expectation on the similarity between the RRG and RP models. This allows us to speak on the special class of models with the explicit (RRG) or hidden (RP) hierarchical structure.

\section{Acknowledgments} We are grateful to B. L. Altshuler, E. Bogomolny, G. Biroli, M. S. Foster, L. B. Ioffe, J. P. Pekola and O. Yevtushenko for stimulating discussions. We would like to specially thank Pragya Shukla for letting us know about her forthcoming work \cite{Shukla}, where the second transition at $\gamma=1$ was mentioned. I.~M.~K. acknowledges the support of the European Union Seventh Framework Programme INFERNOS (FP7/2007 - 2013) under grant agreement no. 308850 and of Academy of Finland 
(Project Nos. 284594, 
272218). 
V.~E.~K. acknowledges the hospitality
of Aalto University and at LPTMS of University of Paris Sud at Orsay and support under the CNRS grant
ANR-11-IDEX-0003-02 Labex PALM, project MultiScreenGlass.

\newpage
\begin{widetext}
\appendix

\section{Numerical extrapolation of the spectrum of fractal dimensions. }\label{SMsec:f(a)}
We start with the computations of the distribution function $P(x)$ of the normalized amplitude $x=N|\psi|^{2}$ in the Rosenzweig-Porter (RP) ensemble and extracted the spectrum of fractal dimensions $f(\alpha,N)=\ln(N\, P_{env}(\ln N (1-\alpha)))/\ln N$, using the approach of Ref.~\cite{Our-BL}. In order to eliminate the effect of zeros of wave functions $\psi$ which dominate the distribution function $P(x)$ at small $x<x_{min}\sim N^{1-\gamma}$ and extract the distribution function of a smooth envelope of $|\psi|^{2}$ we represent $\psi = \psi_{env}\times\eta$, where $\eta$ is the Gaussian Orthogonal Ensemble (GOE) random oscillations with the unit average square which are supposed to be statistically independent of $\psi_{env}$. Then the distribution of $\ln x$ is a convolution of the distribution $P_{env}(y)$ of $y=\ln x_{env}=\ln (N \psi_{env}^{2})$ and the known GOE distribution of $\ln \eta^{2}$. Making a numerical de-convolution one obtains the distribution $P_{env}(y)$ in which the effect of zeros of $\eta(r_{o})$ is eliminated. Such a ``rectified'' distribution function decreases much faster at small $x_{env}<x_{min}$ than the distribution $P(\ln x)$. This ``rectification procedure'' results in a sharp cut-off of $f(\alpha,N)$ at large $\alpha$. Then the so obtained $f(\alpha,N)$ is extrapolated to $N=\infty$ using the ansatz \cite{Our-BL} $f(\alpha,N)=f(\alpha)+ c(\alpha)/\ln N$.

The results for the localized case $\gamma=3$ and $N=2^{8}...2^{14}$ are shown in Fig.~\ref{Fig:falpha300}.
One can see that $f(\alpha)$ rectified and extrapolated as explained above has a linear in $\alpha$ part which exactly coincides with the prediction of the paper (see Fig.~1(a) of the paper). Moreover, the singular peak predicted for $\gamma>2$ in the paper is observed and its top value tends to zero as $N\to\infty$ (see inset in Fig.~\ref{Fig:falpha300}). In Fig.~\ref{Fig:falpha150} we present the extrapolated $f(\alpha)$ for $\gamma=1.5$. Again, the linear part of $f(\alpha)$ coincides with the expectation shown in Fig.~1(c) of the paper.

Finally, we present the results for $f(\alpha,N)$ for $\gamma=0.75<1$ (see Fig.~\ref{Fig:falpha075}). One can see that the distribution function $P(x)$ is weakly $N$-dependent and is almost indistinguishable on the logarithmic scale from the GOE distribution function. The function $f(\alpha,N)$ is more sensitive to $N$ but also in this case there is a minor difference from the corresponding GOE result.

\section{Moments of $|\psi|^{2}$.}\label{SMsec:moments}
Here we discuss the moments
\be
I_{q}=\left\langle \sum_{r}|\psi(r)|^{2q}\right\rangle = N\langle |\psi|^{2q} \rangle \propto N^{-\tau(q)}.
\ee
We start by considering the localized phase $\gamma>2$ where the moments with $q\geq 1$ are dominated by the singular term
$N^{-1} \delta(x-N)$ in the distribution function. One can easily see that all such moments are $N$-independent, which corresponds to $\tau(q)=0$. Using this equality and given that $\tau(q)$ and $f(\alpha)$ are related by the Legendre transform \cite{MirRev}:
\be\label{Legendre}
\tau(q)=q\alpha_{q} - f(\alpha_{q}),\;\;\;\;f'(\alpha_{q})=q,
\ee
and that the singular peak is located at $\alpha=0$, one immediately obtains that the top of the peak corresponds to $f(0)=0$. Considering also $q<1$ and using the Legendre transform Eq.~(\ref{Legendre}) we obtain from $f(\alpha)$ of Fig.~1 of the paper:
\be\label{tau-loc}
\tau(q)=\left\{ \begin{matrix}\gamma q-1, & q<1/\gamma \cr
0, & q>1/\gamma\end{matrix}\right.,\;\;\;\;\;(\gamma>2).
\ee
By the nature of the Legendre transform the tangential to $f(\alpha)$ with any slope $q$ is the same for the singular and ``non-convex'' $f(\alpha)$
shown by the red solid line in Fig.~1(a) of the paper and for the ``convex'' $f(\alpha)$ shown by the blue dashed line.
Thus for both types of $f(\alpha)$ shown in Fig.~1(a) of the paper $\tau(q)$ is given by Eq.~(\ref{tau-loc}).

For $1<\gamma<2$ Eq.~(\ref{Legendre}) gives:
\be\label{ttau}
\tau(q)=\left\{ \begin{matrix}\gamma q-1, & q<1/2 \cr
(2-\gamma)(q-1), & q>1/2\end{matrix}\right.,\;\;\;\;(1<\gamma<2).
\ee
Eq.~(\ref{ttau}) implies that in our model the multifractal dimensions $D_{q}$ for $1<\lambda<2$ do not depend on $q$ for $q>1/2$ and
are equal to:
\be\label{fractal-dim}
D_{q}\equiv\tau(q)/(q-1)=(2-\gamma), \;\;\;\;\;(q>1/2).
\ee

\section{Support set dimension and the curvature.}\label{SMsec:support_set}
We computed numerically the moment $\langle x\,\ln x\rangle$, where $x=N\,|\psi|^{2}$, which in a pure multifractal state should behave as:
\be\label{lin}
\langle x\,\ln x\rangle = (1-D_{1})\,\ln N + const, \;\;\;\; (N\to\infty),
\ee
where $D_{1}$ is a dimension of the wavefunction support set. In an ergodic phase $D_{1}=1$, while in the localized phase $D_{1}=0$. In the intermediate multifractal state $0< D_{1}<1$. The support set dimension \cite{KravAltScard} may be expressed through the solution $\alpha=\alpha_{1}$ of the equation:
\be
f'(\alpha)=1,
\ee
(with $f(\alpha)$ being a spectrum of fractal dimensions) i.e. $D_{1}=\alpha_{1}$ is a point $\alpha=\alpha_{1}$ where the line with the slope $1$ is tangential to $f(\alpha)$. One can immediately calculate $D_{1}$ as a function of $\gamma$ using Fig.~1 of the paper:
\be\label{D1-calc}
D_{1}= \left\{\begin{matrix}1,& (\gamma<1)\cr
2-\gamma, & (1 <\gamma <2)\cr
0,& (\gamma>2)\end{matrix}\right.
\ee

The dependence of the moments $I_{1}=\langle x\,\ln x \rangle$ vs. $\ln N$ is shown in the inset of Fig.~\ref{Fig:moments} in the main text for $N=2^{8},2^{9}...,2^{15}$. By fitting to Eq.~(\ref{lin}) one can find $D_{1}$ and compare it with the expected result Eq.~(\ref{D1-calc}). The comparison is shown in Fig.~\ref{Fig:moments} in the main text. One can see that the apparent $D_{1}$ extracted from limited sizes $N=2^{8}-2^{15}$ deviates from the prediction Eq.~(\ref{D1-calc}) as $\gamma$ approaches the Anderson transition point $\gamma=2$.
We believe that this is a finite-size effect related to the correlation volume $N_{c}\sim\xi^{3}$ that diverges exponentially at the transition. Clearly, for $N<N_{c}$, one should see the properties of the critical point where $D_{1}=0$. That is why the apparent $D_{1}$ is smaller as the prediction in the vicinity of $\gamma=2$. At the same time, Eq.~(\ref{D1-calc}) describes very well the data points for $D_{1}$ close to $\gamma=1$. This probably means that the ergodic transition is not associated with an exponentially divergent correlation length $\xi$.

One can quantify the deviations from the linear behavior Eq.~(\ref{lin}) by introducing the $1/N$ correction which has a finite curvature in the $\ln N$ variable. In reality, the finite-size scaling exponents are not known for this model, and the true corrections could be very different from $1/N$. However,
coefficient $C_{1/N}$ in the simple fit:
\be\label{fit-1N}
\langle x\,\ln x\rangle = (1-D_{1})\,\ln N + c_{0}+ C_{1/N}\,N^{-1}
\ee
gives an idea about the ``global curvature'' of the dependence $\langle x\,\ln x \rangle$ vs. $\ln N$. The dependence of this coefficient on $\gamma$ is shown in Fig.~\ref{Fig:moments} in the main text. It has a characteristic peak shape. For $\gamma$ in the vicinity of $1$ the finite-size effects are small at our system sizes, and the global curvature is small too. Close to the point $\gamma=2$ our system sizes are far too small to deviate from the critical behavior at the transition point. In this case the global curvature is small too. The absolute value of the global curvature reaches its maximum where $\ln N\sim \ln N_{c}$.

An important observation is that the ``global curvature'' changes sign at the ergodic transition. This observation may be used to locate the transition point in finite-size calculations.

In addition to the global curvature one may also introduce a ``local curvatures'' as follows:
\be\label{lcurvature}
\kappa_{<} = \frac{(I_{1}[[1]]+I_{1}[[3]]-2I_{1}[[2]])/\ln^{2}2}{\left( 1+ (I_{1}[[1]]-I_{1}[[3]])^{2}/4\ln^{2}2\right)^{3/2}},\;\;\;
\kappa_{>} = \frac{(I_{1}[[-1]]+I_{1}[[-3]]-2I_{1}[[-2]])/\ln^{2}2}{\left( 1+ (I_{1}[[-1]]-I_{1}[[-3]])^{2}/4\ln^{2}2\right)^{3/2}},
\ee
where $I_{1}[[i]]$ is a moment $I_{1}$ for the $i$-th system size counted from the first one $N=2^{8}$, while $I_{1}[[-i]]$ is the moment $I_{1}$ for the $i$-th system size counted from the last one $N=2^{15}$. Eq.~(\ref{lcurvature}) is nothing but the discrete variant of the curvature:
\be
\kappa = \frac{h''(x)}{(1+[h'(x)]^{2})^{3/2}}
\ee
of a curve given by a function $h(x)$.

In the table below we present $\kappa_{<}$ and $\kappa_{>}$ for different $\gamma$.
\begin{table}[h]
\caption { Local curvatures $\kappa_{<}$ and $\kappa_{>}$.} \label{tab:T1}
\begin{center}
\begin{tabular}{| l || l | c | r | r |}
 \hline
 & $\kappa_{<}\times 10$ & $\kappa_{>}\times 10$ \\ \hline
 $\gamma=0.75$ & $+0.05$ & $+0.05$ \\ \hline
 $\gamma=1.00$ & $-0.16$ & $0.03$ \\ \hline
 $\gamma=1.25$ & -0.42 & -0.10 \\ \hline
 $\gamma=1.40$ & -0.40 & -0.27 \\ \hline
 $\gamma=1.60$ & -0.13 & -0.17 \\ \hline
 $\gamma=1.75$ & -0.03 & -0.06 \\ \hline
 $\gamma=2.00$ & 0.00 & 0.00 \\ \hline
 $\gamma=2.25$ & 0.00 & 0.00 \\
 \hline
\end{tabular}
\end{center}
\end{table}
One can see that the local, as well as the global curvature is negative for $1<\gamma<2$. It is positive and small for $\gamma=0.75 <1$. Thus it is likely, that the local curvature, too, changes sign close to the ergodic transition at $\gamma=1$.
The absolute value of $\kappa$ decrease with increasing $N$ in the vicinity of $\gamma=1$ (for $\gamma=1.25$ and $\gamma=1.40$), which signals about convergence. In contrast $|\kappa|$ increase with increasing $N$ in the vicinity of $\gamma=2$ (for $\gamma=1.6$ and $\gamma=1.75$) as the system size $N$ approaches the correlation volume from below. For $\gamma=2.0$ and $\gamma=2.25$ the local curvature is very small and is inside the error bar.


\section{Overlap correlation function and the Thouless energy.}\label{SMsec:overlap}
\begin{figure}[b]
 \center{\includegraphics[width=0.38\linewidth]{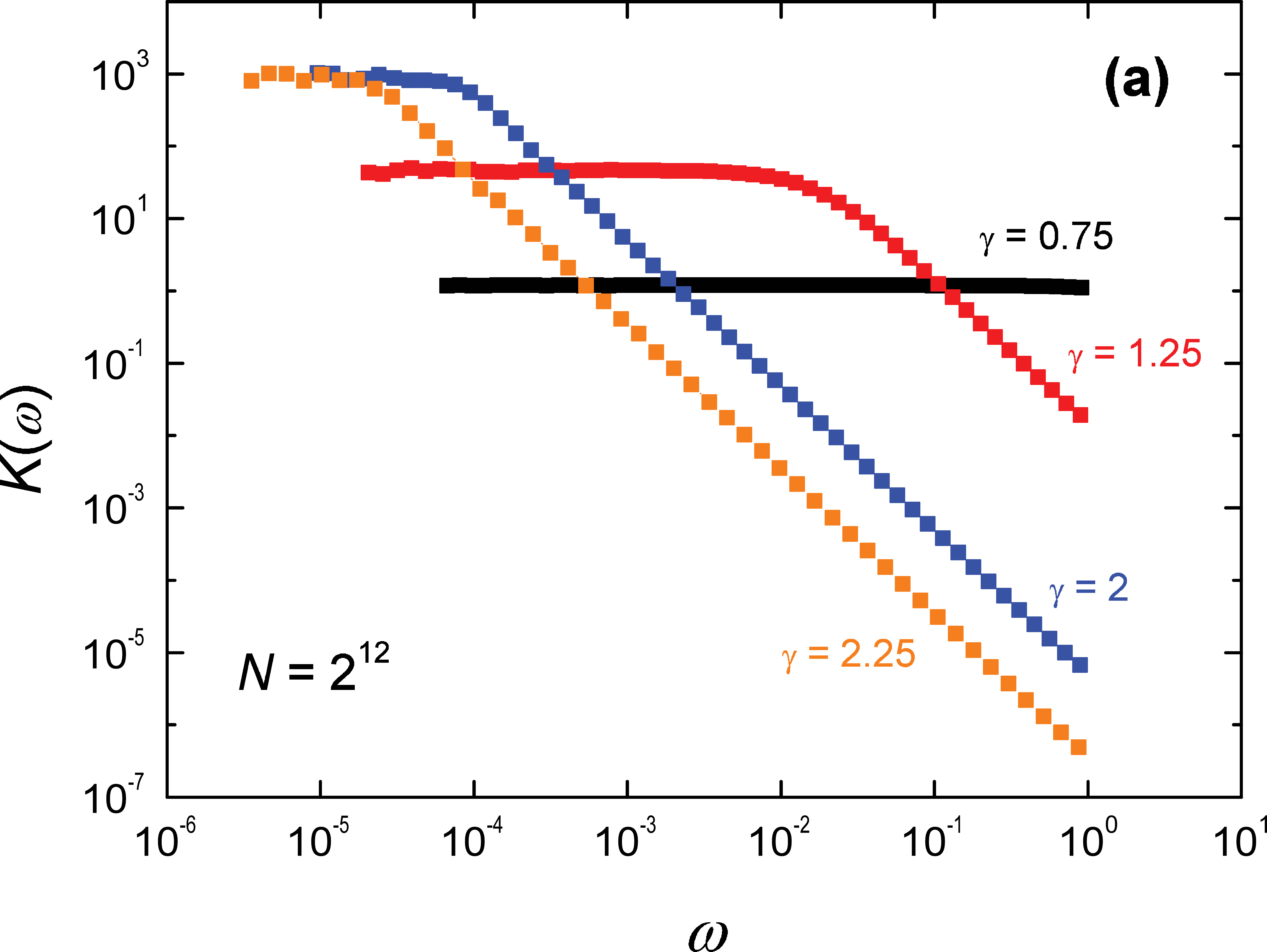}\includegraphics[width=0.37\linewidth]{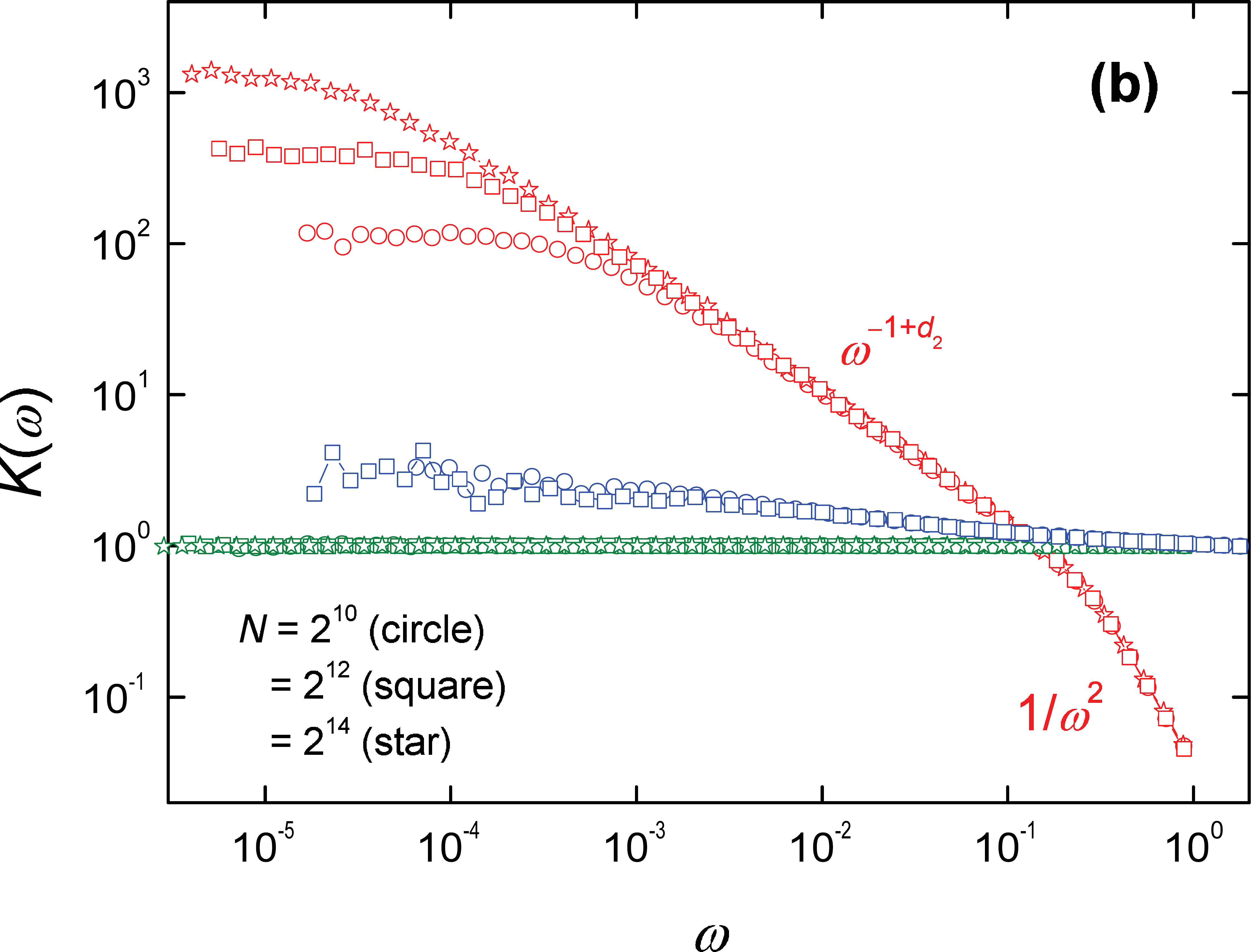}}
 \caption{(Color online) Overlap correlation function Eq.~(\ref{overlap}): (a) for the Rosenzweig-Porter model at different $\gamma$ at $N=2^{12}$; (b) for the PLBRM model (red symbols),GOE (green symbols) and banded random matrix model (blue symbols). There are three regions with different behavior for the PLBRM: the plateau, the Chalker's scaling with the non-trivial exponent $\omega^{-1+D_{2}}$, the fast decay $\sim \omega^{-2}$. Only two of these three regimes are present on the plot (a): the Chalker's scaling is absent. The GOE behavior $K(\omega)=const$ is identical to the one for RP model at $\gamma=0.75$. Note that the behavior in the localized region $\gamma>2$ of the RP model is qualitatively different from that of the quasi-one dimensional localization in the banded random matrices: in the former case the repulsion of wave functions is present, while in the latter case positions of centers of localization are randomly distributed in space with almost no correlations.
 } \label{Fig:Kw}
 \end{figure}
 Conventional multifractal correlations imply not only a power-law scaling of the moments of $|\psi_{n}(r_{o})|^{2}$ with the system size $N$ but also a specific power-law two- and multi- point correlations. In particular, the overlap correlation function:
 \be\label{overlap}
 K(\omega)=N \sum_{r}|\psi_{E}(r)|^{2}\,|\psi_{E+\omega}(r)|^{2}\sim \left(\frac{E_{0}}{\omega} \right)^{\mu},\;\;\;\;\mu=(1-D_{2}),
 \ee
 for $\delta<\omega<E_{0}$ (with $\delta$ being the mean level spacing and $E_{0}\sim O(1)$ being the onset of the anti-correlations) obeys the Chalker's scaling \cite{Chalk-Daniel,KrMut,CueKrav-orig} in the energy domain. Eq.~(\ref{overlap}) holds exactly at the critical point of the 3D Anderson model and in certain random matrix ensembles, e.g. for the power law banded random matrices (PLBRM) \cite{KrMut,Seligman}. It implies an enhancement of correlations compared to the case of independently fluctuating wave function which would result in $K(\omega)=1$. However, Eq.~(\ref{overlap}) is also approximately valid in the metallic phase of the 3D Anderson model close to the transition point \cite{CueKrav-orig} in the limited range of $\delta_{\xi}<\omega<E_{0}$, where $\delta_{\xi}=(\rho N_{c})^{-1}$ is the mean level spacing in the correlation volume $N_{c}\sim\xi^{3}$. For $\omega<\delta_{\xi}$ the correlation function saturates
 developing a plateau, and for $\omega>E_{0}$ it decreases as fast as $\omega^{-2}$ \cite{CueKrav-orig}. In the 3D Anderson model the plateau survives the thermodynamic limit $N\to\infty$ and extends to larger $\omega$ as one goes deeply into the metallic phase. It is thus a signature of the ergodic extended state. In the region $\omega>E_{0}$, the overlap correlation function is small, which signals on the ``repulsion of wave functions'' at large energy separations \cite{CueKrav-orig}. A similar phenomenon of eigenfunction repulsion for $\omega>E_{0}$ was observed in the PLBRM with a small bandwidth $b$ (see \cite{CueKrav-orig} or Fig.~\ref{Fig:Kw}(b)).

 We calculated $K(\omega)$ in our model numerically. The result is presented in Fig.~3 of the paper and in Fig.~\ref{Fig:Kw}(a) in the SM. Surprisingly, no Chalker's scaling and enhancement of correlations similar to Eq.~(\ref{overlap}) was observed. For $1<\gamma<2$ the correlations are fast decreasing at $\omega>E_{Th}\sim N^{1-\gamma}$:
 \be\label{K-intermed}
 K(\omega)\sim \left\{\begin{matrix} N^{1-\gamma}\,\omega^{-2}, & \omega>E_{Th} \cr
 N^{\gamma-1}, & \omega<E_{Th}\end{matrix}\right.,\;\;\;\;\;(1 <\gamma <2)
 \ee
 like in the high-energy region $\omega>E_{0}$ of the 3D Anderson and PLBRM models. The GOE-like plateau is present only in a narrow interval of small $\omega<E_{Th}\sim 1/N^{\gamma-1}$ which shrinks to zero in the thermodynamic limit.
 Thus we can identify $E_{Th}\sim N^{1-\gamma}$ as the {\it Thouless energy} for this model, which by definition is the border line between the GOE-like behavior (plateau) for $\omega<E_{Th}$ and the system-specific behavior for $\omega>E_{Th}$ (fast decay $\propto \omega^{-2}$).
\begin{figure}[b]
 \center{\includegraphics[width=0.36\linewidth]{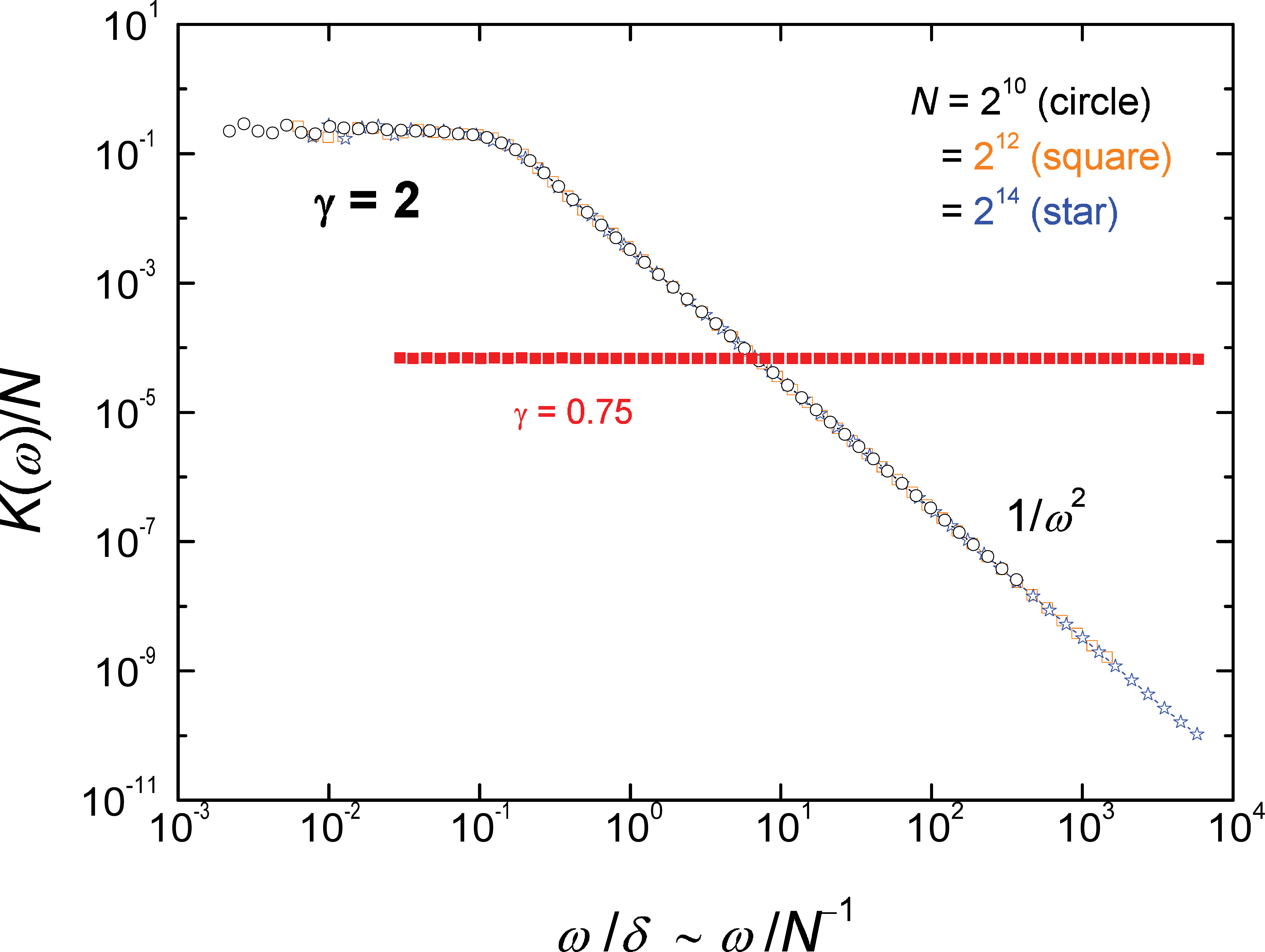}\includegraphics[width=0.40\linewidth]{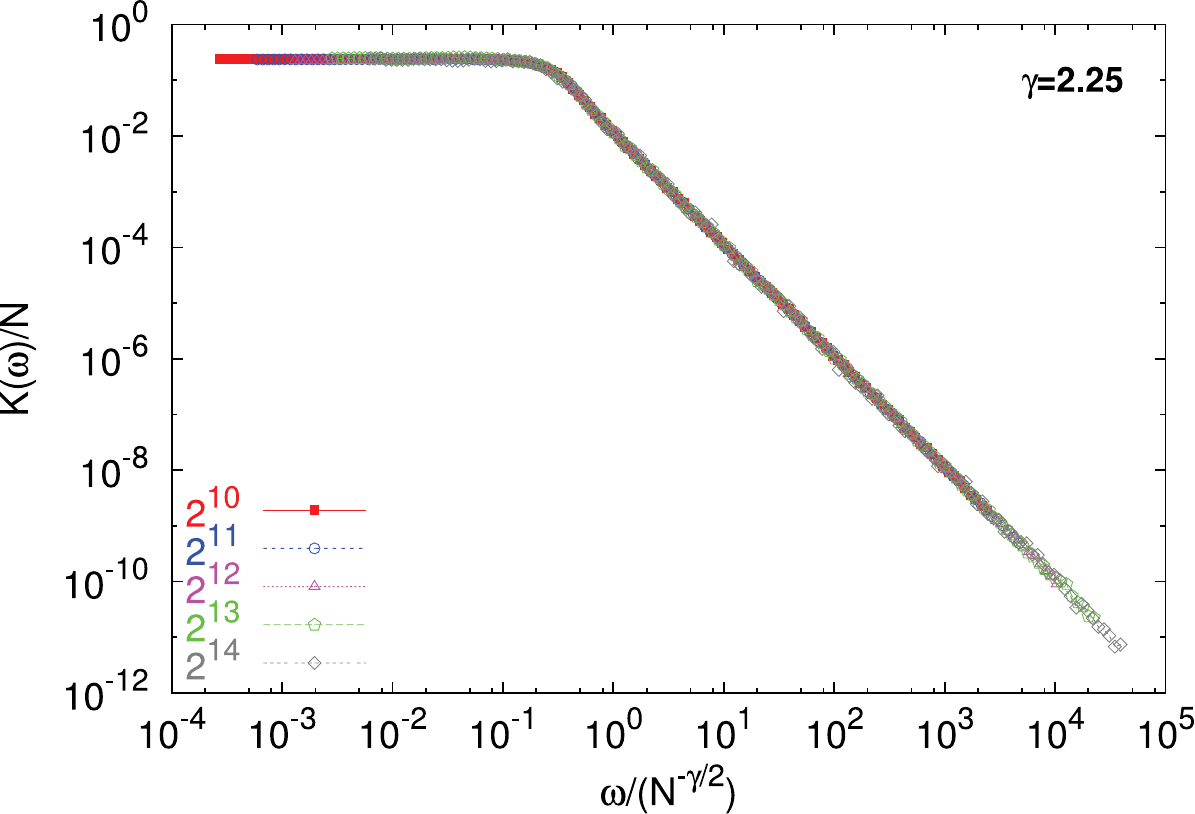}}
 \caption{(Color online) Collapse of data for $K(\omega)$ at $\gamma=2.00$ and $\gamma=2.25$ in the coordinates $X=\omega/N^{-\gamma/2}$, $Y=K(\omega)/N$.
 } \label{Fig:collapse-250}
 \end{figure}

 Note that the onset of the anti-correlations scales with $N$ exactly as the parameter $E_{Th}=\delta N^{2-\gamma}$ which enters the dimensionless variable $u=\frac{1}{2}(t-t')\,E_{Th}$ in the unfolded spectral form-factor $S(u)$ in the level statistics (shown in the inset of Fig.~\ref{Fig:R(u)} in the main text). It can be nicely expressed in terms of the number of populated sites $N^{D_{1}}$ in the wave function support set:
 \be\label{Thouless}
 E_{Th}=\delta\,N^{D_{1}},\;\;\;\;(1<\gamma<2).
 \ee

 Qualitatively similar (but quantitatively different) behavior is observed in the localized phase $\gamma>2$:

 \be\label{K-loc}
 K(\omega)\sim \left\{\begin{matrix} N^{1-\gamma}\,\omega^{-2}, & \omega>E_{Th}\sim N^{-\gamma/2} \cr
 N, & \omega<E_{Th} \end{matrix}\right.,\;\;\;\;\;\;(\gamma>2)
 \ee
 In this case we have:
 \be\label{Thouless2}
 E_{Th}\sim N^{-\gamma/2},\;\;\;\;\; (\gamma\geq 2).
 \ee
Surprisingly at $\gamma>1$ and a fixed $\omega$ the overlap $K(\omega)$ drops below the independent wave function limit $K(\omega)=1$. Thus in $N\to\infty$ limit of our model for $\gamma>1$ the repulsion of wave functions happens at all energy scales.
\begin{figure}[h]
 \center{\includegraphics[width=0.38\linewidth]{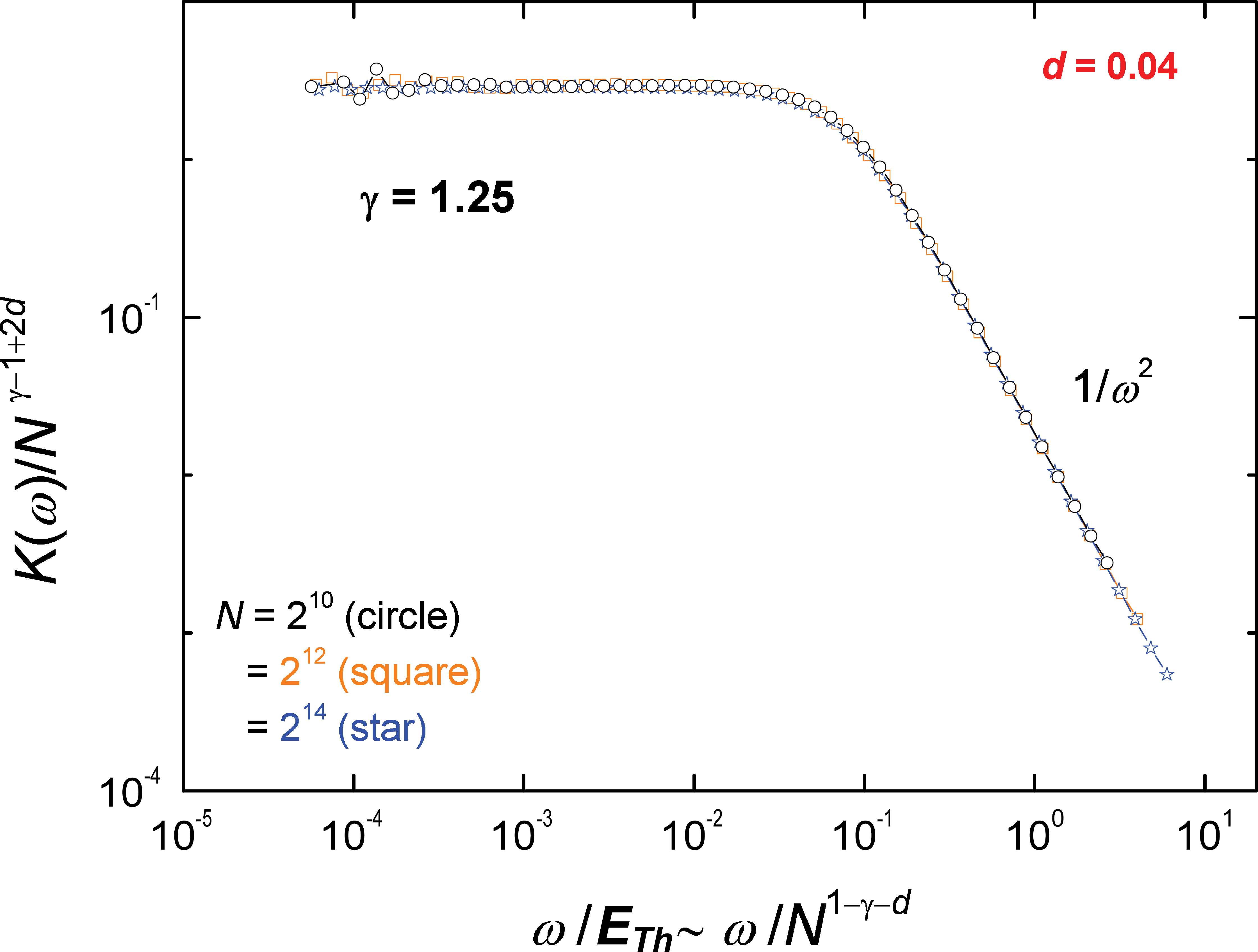}\includegraphics[width=0.37\linewidth]{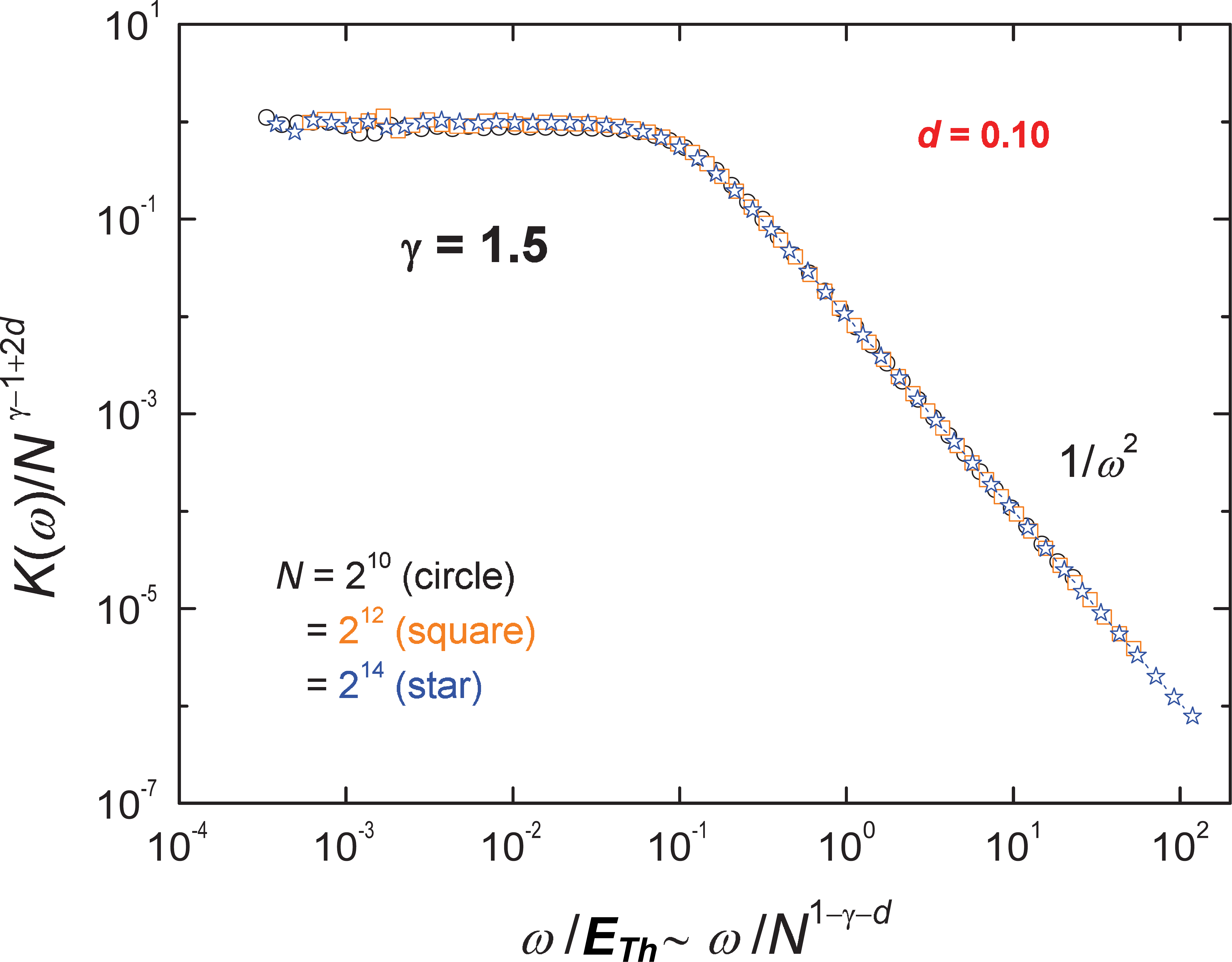}}
 \caption{(Color online) Collapse of data for $K(\omega)$ at $\gamma=1.25$ and $\gamma=1.50$ in the coordinates $X=\omega/N^{1-\gamma-d}$, $Y=K(\omega) N^{1-\gamma-2d}$. The corrections $d=0.04$ and $d=0.10$ are obtained from the best collapse.
 } \label{Fig:collapse-d}
 \end{figure}
 \begin{figure}[h]
 \center{\includegraphics[width=0.75\linewidth]{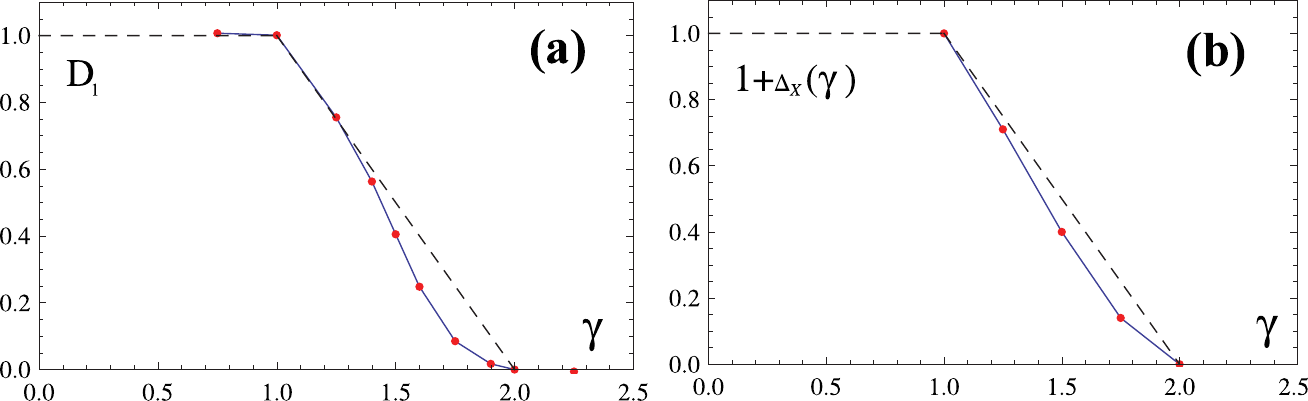}}
 \caption{(Color online) (a) The support set dimension $D_{1}$ vs. $\gamma$. (b) The scaling exponent $\Delta_X(\gamma)$ vs. $\gamma$. In both cases the finite-size effects make $D_{1}$ and $1+\Delta_X$ closer to its value at $\gamma=2$ than expected (dashed line).
 } \label{Fig:d-D}
 \end{figure}

 For $\gamma<1$ the plateau $\sim 1$ in $K(\omega)$ extends to almost the entire spectral bandwidth (see Fig.~\ref{Fig:Kw}(a)) and may even exceed it:
 \be\label{Thouless-erg}
 E_{Th}\sim O(N^{0}),\;\;\;\;\;(\gamma\leq 1)
 \ee
 The emergence at $\gamma=1$ of a plateau that survives the thermodynamic limit and occupies a finite fraction of (or all of) the spectral bandwidth is a very clear signature of the ergodic transition at $\gamma=1$.

Eqs.~(\ref{K-intermed}, \ref{K-loc}) were checked by a collapse of the data points for a fixed $\gamma$ but different $N=2^{10}- 2^{14}$. The results are shown in Fig.~\ref{Fig:overlap} in the main text and Fig.~\ref{Fig:collapse-250}.

As $\gamma$ approaches the Anderson transition point $\gamma=2$ deviations from Eq.~(\ref{K-intermed}) emerge. For $1<\gamma<2$ the best collapse was found to occur in the coordinates $X=\omega N^{\Delta_X(\gamma)}$, $Y=K(\omega)/N^{\Delta_Y(\gamma)}$, where:
\be\label{a-b-inter}
\Delta_X(\gamma)=\gamma-1+d,\;\;\;\;\Delta_Y(\gamma)=\gamma-1-2d.
\ee
The correction $d(\gamma)$ is most probably due to the finite-size $N$ that could be comparable with the correlation volume near the Anderson transition point $\gamma=2$. The dependence $1+\Delta_X(\gamma)$ vs. $\gamma$ is plotted in Fig.~\ref{Fig:d-D}(b). It has an apparent analogy with the dependence $D_{1}(\gamma)$ of the support set dimension in Fig.~\ref{Fig:d-D}(a). We believe that in both cases this is a finite-size effect when $N\leq e^{const/(2-\gamma)}$
is smaller than the exponentially large correlation radius near the AT point $\gamma=2$.

\section{The spectral form-factor.}\label{SMsec:form-factor}
Here we consider the spectral form-factor $C(t,t')=\sum_{n\neq m}e^{it E_{m}+it' E_{n}}$ ($\{E_{n}\}$ is a set of eigenvalues of $H$) which was derived for URP model Eq.~(1) in the main text using the Itzikson-Zuber formula of integration over unitary group and we start by Eq.~(3.2) of Ref.~\cite{ShapiroKunz}
\be \label{form-fac}
C(t,t') = \frac{e^{-\sigma (t^2+t'^2)/2}}{N\tau\tau'}\oint_{\Gamma_R} \frac{dz}{2\pi i}\oint_{\Gamma_R} \frac{dz'}{2\pi i}e^{i (t z+ t' z')}
\left[g(z,z')^N\left(1+\frac{\tau \tau'}{(z'-z-\tau)(z'-z+\tau')}\right)-\rho(z)^N\rho(z')^N\right] \ .
\ee
In this exact formula integration is extended over the contour $\Gamma_{R}$ that encompasses the real axis and $\tau=it\sigma$. The functions $g(z,z')$ and
$\rho(z)$ depend on the diagonal disorder distribution $p(a)$ through $\alpha(z)=\int {p(a)da}/{(z-a)}=\langle (z-a)^{-1}\rangle$ and are defined as follows: $\rho(z)=1+\tau\alpha(z)$,
\be\label{g-rho}
 g(z,z')=1+\tau\alpha(z)+\tau'\alpha(z')-\frac{\tau\tau'}{z-z'}\,[\alpha(z)-\alpha(z')].
\ee
In order to do the limit of infinite matrix size $N\to\infty$ we make a re-scaling:
\bea\label{rescaling1}
(t,t')&=&\frac{T}{2}\pm N^{\Delta_{t}}\,s\\
(z,z')&=& x\pm\frac{y}{2N^{\Delta_{z}}}-\frac{i\epsilon}{N^{\Delta_{z}}}\,(q,q')\label{rescaling2},
\eea
where $\epsilon\to +0$ and $q,q'=\pm 1$ for the part of the contour below and above the real axis. The exponents $\Delta_{t}>0$ and $\Delta_{z}>0$ should be chosen so that
(i) finite limits $\lim_{N\to\infty} [g(z,z')]^{N}$ and $\lim_{N\to\infty} [\rho(z)]^{N}$ exist, and
(ii) the entire expression $C(t,t')$ is finite in the $N\to\infty$ limit.
One can show that the choice
\be\label{rescaling-Delta}
\Delta_{t}=\gamma-1,\;\;\;\;\Delta_{z}=1,
\ee
satisfies all these conditions if $\gamma>1$. Now doing the limit $N\to\infty$ at fixed $T,s,x,y$ we observe that the parameter $\gamma$ enters the limiting expression Eq.~(\ref{form-fac}) only in the exponents $e^{-\sigma (t^2+t'^2)/2}\to e^{-\lambda^{2}N^{\gamma-2}\,s^{2}}$ and $e^{i (t z+ t' z')}=e^{iT x+i N^{\gamma-2} s y}$.

This observation immediately tells us that for $\gamma>2$ the spectral form-factor $C(t,t')$ vanishes, which leads to completely uncorrelated energy levels and the exact Poisson statistics. In the case $\gamma=2$ considered in Ref.~\cite{ShapiroKunz} the level statistics is different from Poisson, as $C(t,t')=2\pi p(0)\delta(T)\,[S(u=s/p(0))-1]$ is not zero. However, the level compressibility equals unity
\be\label{chi}
\chi=S(0)=1.
\ee
like for uncorrelated energy levels.

One can easily see that Eq.~(\ref{chi}) remains valid also in the entire region $1<\gamma\leq 2$, since at $s=0$ the $\gamma$-dependent exponents $e^{-\lambda^{2}N^{\gamma-2}\,s^{2}}$ and $ e^{i \,N^{\gamma-2}\,s y}$ are equal to 1 anyway.

Eq.~(\ref{form-fac}) can be cast as follows $C(t,t')=e^{-\sigma (t^2+t'^2)/2}\,(K_{1}(t,t')+K_{2}(t,t'))$, where:
 \begin{gather}
K_1(t,t') = \frac{1}{N\tau\tau'}\oint_{\Gamma_R} \frac{dz}{2\pi i}\oint_{\Gamma_R} \frac{dz'}{2\pi i}e^{i (t z+ t' z')}
\left(g(z,z')^N-\rho(z)^N\rho(z')^N\right)
 \ ,
\end{gather}
\begin{gather}
K_2(t,t') = \frac{1}{N}\oint_{\Gamma_R} \frac{dz}{2\pi i}\oint_{\Gamma_R} \frac{dz'}{2\pi i}e^{i (t z+ t' z')}
\frac{g(z,z')^N}{(z'-z-\tau)(z'-z+\tau')} \ .
\end{gather}

Performing the re-scaling Eqs.~(\ref{rescaling1}~--~\ref{rescaling-Delta}) of the paper and changing the variables $y\to (q-q') y$ we arrive at:
\be\label{K_1}
K_1(t,t') =
\sum_{q}\int \frac{p(x)dx}{a}e^{i x T-2\pi \lambda^2 s q p(x)}\int \frac{dy}{2\pi}e^{i b(y-i\epsilon)}
\left(e^{-\frac{i a}{y-i\epsilon}}-1\right)
 \ ,
\ee
\be\label{K_2}
K_2(t,t') =
\sum_{q}\int \frac{dx}{2\pi}e^{i x T-2\pi \lambda^2 s q p(x)}\int \frac{dy}{4\pi}e^{i b(y-i \epsilon)}
\frac{e^{-\frac{i a}{y-i \epsilon}}}{(y-i \epsilon+i s q \lambda^2/2)^2}
-\sum_{q}\int \frac{dx}{2\pi}e^{i x T}\int \frac{dy}{4\pi}e^{i b y}
\frac{1}{(y+i s q\lambda^2/2)^2}
 \ ,
\ee
where $p(x)$ is the distribution of $H_{nn}$ coinciding with the density of states for $\gamma>1$ \cite{ShapiroKunz}, the limit $N\to\infty$ is taken everywhere, except in the exponents $e^{i (t z+ t' z')}\to e^{i N^{\gamma-2}s (q-q')(y-i \epsilon)}$, the summation over $q'$ is taken, and:
\be
a=\pi \lambda^4 s^2 p(x),\;\;\;b=2 s q N^{\gamma-2}.
\ee
The exponents determine the allowed contour deformations. Deforming the contour so that the exponents are small at $|y|\to 0$ we observe that both $K_{1}(t,t')$ and $K_{2}(t,t')$ are identically zero for $s q<0$. Thus the summation over $q$ can be dropped with $s$ being replaced by $|s|$. One can also express the integrals over $y$ in terms of the Bessel functions, deforming the contour so that it encompasses the origin along the infinitesimal circle. The final result for $C(t,t')=2\pi \delta(t+t')\,[S(u=s/p(0))-1]$ reads:
\bea\label{SS}
S(u)&=&
 1 + e^{-2 \pi \Lambda^{2} u} \frac{
 2 e^{-\Lambda^{2} u^{2} N^{\gamma - 2}}}{ \sqrt{
 8 \pi N^{\gamma - 2} \Lambda^{4} u^{3}}}
 I_{1}( \sqrt{
 8 \pi N^{\gamma - 2} \Lambda^{4} u^{3}})\\ \nonumber &-&
 \frac{1}{4 \pi} \sqrt{
 8 \pi \Lambda^{4} u^{5}}
 N^{(3 \gamma - 6)/2} e^{-2 \pi \Lambda^{2} u}
 e^{-\Lambda^{2} u^{2} N^{\gamma - 2}}\,\int_{0}^{\infty}\frac{x\,dx}{\sqrt{x+1}}\,I_{1}(\sqrt{
 8 \pi N^{\gamma - 2} \Lambda^{4} u^{3}\,(x+1)})\,e^{-x\,u^{2}\Lambda^{2}N^{\gamma-2}},
\eea
coinciding with Eq.~(9) in the main text.
Here $I_{1}(x)$ is the modified Bessel function and $\Lambda=\lambda p(0)$. Eq.~(\ref{SS}) is valid for $\gamma>1$.
The second and the third terms in Eq.~(\ref{SS}) correspond to $K_{1}(t,t')$ and $K_{2}(t,t')$, respectively. At $\gamma>2$ both of the terms vanish in the $N\to\infty$ limit, and the Poisson limit $S(u)=1$ is reached; for $\gamma=2$ both the terms are non-zero; for $1<\gamma<2$ the third term cancels the first one and the second term tends to a finite limit as $N\to\infty$:
\be\label{limit}
\lim_{N\to\infty}S\left(u=\tfrac{1}{2}(t-t')\delta\,N^{2-\gamma}\right)= e^{-2\pi \Lambda^{2}\,u},
\ee
The unfolded spectrum form factor is given by:
\be
R\left(\mathfrak{t}\equiv\tfrac{1}{2}(t-t')\delta\right)=S\left(u\equiv\mathfrak{t}N^{2-\gamma}\right),
\ee
where $\delta$ is the mean level spacing.

For $\gamma>2$ one can see that $R(\mathfrak{t})$ tends to the Poisson limit $R(\mathfrak{t})=1$ as $N\to\infty$.

For $1<\gamma<2$ and any $\mathfrak{t}>0$ it evolves towards the Gaussian Unitary Ensemble (GUE) form factor:
\be
R_{GUE}(\mathfrak{t})=\left\{\begin{matrix} \mathfrak{t}/(2\pi), & 0\leq\mathfrak{t}<2\pi\cr
1,& \mathfrak{t}\geq 2\pi \end{matrix}\right.,
\ee
as $N\to\infty$.

However, in contrast to $R_{GUE}(\mathfrak{t})$ the function $R(\mathfrak{t})$ has a jump at $\mathfrak{t}=0$:
\be
R(\mathfrak{t}=0)=1.
\ee
On the other hand, the function $S(u)=R(\mathfrak{t}= u N^{\gamma-2})$ at $1<\gamma<2$ has a finite non-singular limit Eq.~(\ref{limit}) as $N\to\infty$:
\be\label{limit1}
\lim_{N\to\infty}R(u N^{\gamma-2} )= e^{-2\pi \Lambda^{2}\,u},
\ee
while for the true GUE form-factor this limit is zero:
\be
\lim_{N\to\infty}R_{GUE}( u N^{\gamma-2} )=0.
\ee
The limit Eqs.~(\ref{limit},\ref{limit1}) and the new emergent energy scale (the Thouless energy)
\be\label{TE}
E_{Th}=\delta\,N^{2-\gamma}\sim N^{1-\gamma}.
\ee
that separates repulsion of energy levels at small energy difference (large ``time'' difference $\mathfrak{t}$) from attraction of energy levels at large energy difference (small $\mathfrak{t}$),
is a hallmark of the extended non-ergodic state for $1<\gamma<2$.

For $\gamma\leq 1$ Eq.~(\ref{SS}) does not hold. The formal reason is that in the re-scaling Eq.~(\ref{rescaling1}) the second term is no longer the leading one as $N\to\infty$. The physical reason is that at $\gamma\leq 1$ the Thouless energy Eq.~(\ref{TE}) is no longer small compared to the spectral band-width $E_{DoS}$ (the width of the mean density of states $\rho(\varepsilon)$). Note that $S(u=\frac{1}{2}(t-t')E_{Th})$, which is the Fourier transform of the two-level correlation function, must saturate when $(t-t')E_{Th}$ gets smaller than $E_{Th}/E_{DoS}$, where
$E_{DoS}$ is the maximal energy difference $\omega$ within the
spectral band-width:
\be\label{Smax}
S(u)= e^{-2\pi \Lambda^{2}\,E_{Th}/E_{DoS}},\;\;\;\;(u\ll E_{Th}/E_{DoS}).
\ee
If $E_{Th}/E_{DoS}$ is formally divergent, as it is the case for $\gamma<1$, we conclude from Eq.~(\ref{Smax}) that $S_{max}\to 0$. This is how the true GOE limit $S(u)=0$ is reached at $\gamma<1$.

Note that the fact that $E_{Th}\gg \delta$ affects the level number variance ${\rm var}(n)$ ($n$ and ${\rm var}(n)$ are the average number of levels and the level number variance in a certain spectral window, respectively) in which a new scale $E_{Th}/\delta \sim N^{2-\gamma}$ appears for $n$ at $1<\gamma<2$:
\be\label{LNV}
{\rm var}(n)=\left\{ \begin{matrix} \sim \ln n, & 1\ll n\ll N^{2-\gamma}\cr
n, & N^{2-\gamma}\ll n \ll N. \end{matrix}\right.
\ee
The level compressibility $\chi=\lim_{N\to \infty,\atop n\to \infty}\frac{{\rm var}(n)}{n}=0$ if $\lim_{N\to \infty,\atop n\to \infty}n/N^{2-\gamma}=0$ and $\chi=1$ if $\lim_{N\to \infty,\atop n\to \infty}n/N^{2-\gamma}=\infty$ but $\lim_{N\to \infty,\atop n\to \infty}n/N =0$.

\end{widetext}


\begin{thebibliography}{99}
\bibitem{BAA} D. M. Basko, I. L. Aleiner and B. L. Altshuler, Annals of Physics {\bf 321} , 1126 (2006).
\bibitem{Huse} V. Oganesyan, D. A. Huse, Phys. Rev. B {\bf 75}, 155111 (2007).
\bibitem{AltshulerJoffe} M. Pino, B. L. Altshuler, L. B. Ioffe, arXiv: 1501.03853.
\bibitem{AltshulerShlyapnikov} V.P. Michal, I.L. Aleiner, B.L. Altshuler, G.V. Shlyapnikov, arXiv: 1502.00282.
\bibitem{AbouChac-And} R. Abou-Chacra, P. W. Anderson and D. J. Thouless, J. Phys. C: Solid State Physics, {\bf 6}, 1734 (1993).
\bibitem{Mir-Fyod-sparse} Y. V. Fyodorov and A. D. Mirlin, Phys. Rev. Lett. {\bf 67}, 2049 (1991); Nucl. Phys. B {\bf 336}, 507 (1991);
A. D. Mirlin and Y. V. Fyodorov Phys. Rev. B {\bf 56} 13393 (1997).
\bibitem{Biroli} G. Biroli, A. Ribeiro-Texiera, M. Tarzia, arXiv:1211.7334.
\bibitem{Our-BL} A. De Luca, B. L. Altshuler, V. E. Kravtsov and A. Scardicchio, Phys. Rev. Lett. {\bf 113} 046806 (2014).
\bibitem{Biroli-Levy} E. Targuini, G. Biroli, M. Tarzia, arXiv:1507.00296.
\bibitem{Mehta} M. L. Mehta, \emph{Random Matrices}, 3rd ed., (Elsevier/Academic Press, Amsterdam,  2004)
\bibitem{Smiliansky} I. Oren, A. Godel and U. Smilansky J. Phys. A: Math. Theor. {\bf 42} 415101 (2009);
I. Oren and U. Smilansky J. Phys. A: Math. Theor. {\bf 43} 225205 (2010).
\bibitem{RPort} N. Rosenzweig and C. E. Porter, Phys. Rev. B {\bf 120}, 1698 (1960).
\bibitem{Pandey} A. Pandey, Chaos Solitons Fractals {\bf 5}, 1275 (1995).
\bibitem{BrezHik} E. Brezin and S. Hikami, Nucl. Phys. B {\bf 479}, 697 (1996).
\bibitem{Altland-Shapiro} A. Altland, M. Janssen and B. Shapiro, Phys. Rev. E {\bf 56}, 1471 (1996).
\bibitem{ShapiroKunz} H. Kunz and B. Shapiro, Phys. Rev. E {\bf 58},400 (1998).
\bibitem{Shukla2000} P. Shukla, Phys. Rev. E {\bf 62}, 2098 (2000).
\bibitem{Shukla2005} P. Shukla, J. Phys: Condens. Matter, {\bf 17}, 1653 (2005).
\bibitem{RMTLect} V. E. Kravtsov, Lectures on Random Matrix Theory, arXiv:0911.0639.
 \bibitem{MirRev} F. Evers and A. D. Mirlin, Rev. Mod. Phys. {\bf 80}, 1355 (2008).
\bibitem{MirFyod} A. D. Mirlin, Y. V. Fyodorov, A. Mildenberger, and F. Evers, Phys. Rev. Lett. {\bf 97}, 046803 (2006).
\bibitem{KravAltScard} A. De Luca, A. Scardicchio, V. E. Kravtsov and B. L. Altshuler, ArXiv:1401.0019.
\bibitem{Fyod-corr} A. D. Mirlin and Y. V. Fyodorov, Phys.Rev. Lett. {\bf 72} 526 (1994);
A. D. Mirlin and Y. V. Fyodorov, J. Phys. I France {\bf 4} 655 (1994).
\bibitem{frozen1} C. C. Chamon, C. Mudry, and X.-G. Wen, Phys. Rev. Lett. {\bf 77},
4194 (1996).
\bibitem{frozen2} S. Ryu and Y. Hatsugai, Phys. Rev. B {\bf 63}, 233307 (2001).
\bibitem{frozen3} D. Carpentier and P. Le Doussal, Phys. Rev. E {\bf 63}, 026110 (2001).
\bibitem{Foster}  Yang-Zhi Chou and M. S. Foster, Phys.  Rev.  B {\bf 89}, 165136 (2014)
\bibitem{Altshuler_Shklovskii} B.~L.~Altshuler and B.~I.~Shklovskii, Zh.~Eksp.~Teor.~Fiz. {\bf 91}, 220 (1986) [Sov. Phys. JETP {\bf 64}, 127 (1986)].
\bibitem{CueKrav-orig} E. Cuevas and V. E. Kravtsov, Phys. Rev. B. {\bf 76}, 235119 (2007).
\bibitem{Seligman} A. D. Mirlin, Y. V. Fyodorov, F. M. Dittes, J. Quezada,
T. H. Seligman, Phys. Rev. E {\bf 54}, 3221 (1996).
\bibitem{Chalk-Daniel} J. T. Chalker, Physica A {\bf 167}, 253 (1990);
J. T. Chalker, G. J. Daniell, Phys. Rev. Lett. {\bf 61}, 593 (1988).
\bibitem{ChalkKravLer} J. T. Chalker, V. E. Kravtsov and I. V. Lerner, JETP Lett., {\bf 64}, 386 (1996) [Pis'ma v ZhETF, {\bf 64}, 355 (1996)].
\bibitem{KrMut}V. E. Kravtsov, K. A. Muttalib, Phys. Rev. Lett. {\bf 79}, 1913 (1997).
\bibitem{KrOsYev}  V. E. Kravtsov, A. Ossipov, O. M. Yevtushenko and E. Cuevas Phys. Rev. B {\bf 82} 161102 (2010).
\bibitem{Shukla}
S. Sadhukhan and P. Shukla J. Phys. A: Math. Theor. {\bf 48} 415002 (2015); {\it ibid.} 415003 (2015).
\end{thebibliography}
\end{document}